%% file: manuscript.tex
\documentclass[a4paper,fleqn,useAMS,usenatbib]{mnras}

\pdfoutput=1

\usepackage{mathptmx}

\usepackage[T1]{fontenc}
\usepackage{ae,aecompl}

\usepackage[dvips]{graphicx}
\usepackage{longtable}
\usepackage{amsmath}
\usepackage{amsfonts}
\usepackage{amssymb}
\usepackage{relsize}
\usepackage{comment}
\usepackage{booktabs}
\usepackage[dvipsnames]{xcolor}
\usepackage[%
        ]{hyperref}
        
\hypersetup{
	colorlinks=true,	
	urlcolor=MidnightBlue,		
	pdfpagelabels=true,
	hypertexnames=true,
	plainpages=false,
	naturalnames=true,
	pdftitle={Kepler's Dark Worlds},    
	pdfauthor={Jansen \& Kipping},     
	linkcolor=WildStrawberry,          
	citecolor=ForestGreen,        
}

\input{shortcuts.tex}

\title[Kepler's dark worlds]{
\kepler's dark worlds:
A low albedo for an ensemble of Neptunian and Terran exoplanets
}
\author[Jansen \& Kipping]{Tiffany Jansen$^{1}$\thanks{E-mail:
\href{mailto:jansent@astro.columbia.edu}{jansent@astro.columbia.edu}} and David Kipping$^{1}$\\
$^{1}$Dept. of Astronomy, Columbia University, 550 W 120th Street, New York NY 10027}

\date{Accepted . Received ; in original form }

\pubyear{2018}

\begin{document}
\label{firstpage}
\pagerange{\pageref{firstpage}--\pageref{lastpage}}
\maketitle

\begin{abstract}
Photometric phase curves provide an important window onto exoplanetary
atmospheres and potentially even their surfaces. With similar amplitudes to
occultations but far longer baselines, they have a higher sensitivity to
planetary photons at the expense of a more challenging data reduction in terms
of long-term stability. In this work, we introduce a novel non-parametric
algorithm dubbed \phasma\ to produce clean, robust exoplanet phase curves and
apply it to 115 Neptunian and 50 Terran exoplanets observed by \kepler. We
stack the signals to further improve signal-to-noise, and measure an average
Neptunian albedo of $A_g < 0.23$ to 95\% confidence, indicating a lack of
bright clouds consistent with theoretical models. Our Terran sample provides
the first constraint on the ensemble albedo of exoplanets which are most likely
solid, constraining $A_g < 0.42$ to 95\% confidence. In agreement with our
constraint on the greenhouse effect, our work implies that \kepler's solid
planets are unlikely to resemble cloudy Venusian analogs, but rather dark
Mercurian rocks.
\end{abstract}


\section{Introduction}
\label{sec:intro}

\input{introduction.tex}

\section{Data Processing}
\label{sec:data}

\input{data.tex}

\section{Forward Model}
\label{sec:model}

\input{model.tex}

\section{Regression}
\label{sec:regression}

\input{regression.tex}

\section{Results}
\label{sec:results}

\input{results.tex}

\section{Discussion}
\label{sec:discussion}

\input{discussion.tex}

\pagebreak

\section*{Acknowledgments}

This research has made use of the \forecaster\ predictive package by
\citet{chen:2017a}, the {\tt corner.py} code and \emcee\ package by Dan Foreman-Mackey at
\href{http://github.com/dfm/corner.py}{github.com/dfm/corner.py} and \href{https://github.com/dfm/emcee}{github.com/dfm/emcee}, \scipy, \astropy, and the NASA
Exoplanet Archive, which is operated by the California Institute of Technology,
under contract with the National Aeronautics and Space Administration under the
Exoplanet Exploration Program.

Thanks to David Armstrong for sending us his team's HAT-P-7b detrended phase
curve, and Brian Jackson for sharing his {\tt IDL} code {\tt alpha\_beam}. We would also like to thank the reviewer for their helpful comments.

\input{longtables.tex}

\bsp
\label{lastpage}
\end{document}

%% file: shortcuts.tex
\newcommand{\kepler}{\emph{Kepler}}

\newcommand{\python}{{\tt PYTHON}}
\newcommand{\phasma}{{\tt phasma}}
\newcommand{\forecaster}{{\tt forecaster}}
\newcommand{\emcee}{{\tt emcee}}

\newcommand{\scipy}{{\tt scipy}}
\newcommand{\astropy}{{\tt astropy}}
\newcommand{\sinc}{\mathrm{sinc}}

%% file: introduction.tex
The amount of light reflected and emitted from an exoplanet is a direct probe of said planet's surface or atmospheric composition. Several observational strategies have been successful in detecting this light, including high-dispersion Doppler spectroscopy \citep{snellen:2010,birkby:2013}, occultation detections both photometrically \citep{deming:2005} and spectroscopically \citep{charbonneau:2008}, and photometric phase curves \citep{knutson:2007}. The study of these planetary photons, particularly when coupled with transit spectroscopy \citep{seager:2000,charbonneau:2002}, has been instrumental in shaping our understanding of exoplanetary atmospheres to date \citep{burrows:2014}.

In cases where these planetary photons are detected and the planet is known to be a transiting body, it is usually possible to measure the planetary albedo. This is achieved by noting that the occultation and phase curve amplitude is approximately equal to the geometric albedo multiplied by $(R_P/a)^2$ \citep{winn:2010}, and the planetary radius and semi-major axis in units of the stellar radius ($p$ and $a_R$) can be measured using the transit itself \citep{seager:2003}. Contextually, measuring an albedo close to unity implies that the planet is more likely to have an abundance of highly reflective cloud coverage or have an icy surface, rather than, say, a bare surface darkened by basaltic rock \citep{hu:2012}. The shape of a planet's phase curve can also provide insight into a planet's thermal properties; for example an offset of the peak from the point of occultation implies a certain degree of heat redistribution from the substellar point \citep{knutson:2007} or asymmetric reflection from patchy clouds \citep{demory:2013}. In any case, measuring the amplitude and shape of phase curves allows us to identify individual planetary characteristics, which furthers our understanding of planetary compositions on a broader scale.

Amongst the various means of detecting planetary photons, photometric occultations and phase curves are particularly attractive from a data perspective, since transit survey missions, such as \kepler, collect long baselines of thousands of planetary systems at high precision. Further, although both the phase curve and occultation effects have similar amplitudes, the signal to noise ratio (SNR) is actually far superior using the phase curve method. This is because the phase curve temporal baseline is increased by the ratio of the orbital period to the occultation duration, which in turn should improve the SNR as ${\sim}t^{1/2}$. As an example of this, the amplitude of TrES-2b was first detected using just five months of \kepler\ data from the phase curve \citep{kipping:2011}, giving $(6.5\pm1.9)$\, parts per million (ppm), before being later found in occultation by collating 2.7 years of \kepler\ data \citep{barclay:2012}, giving $(6.5\pm1.8)$\,ppm.

As eluded to, \kepler\ data has already been used to detect numerous individual optical phase curves \citep{borucki:2009,kipping:2011,esteves:2013,angerhausen:2015}, although primarily for Jupiter-sized planets on short-periods, such as HAT-P-7b. This is largely as a result of the strong detection bias intrinsic to the method itself, with the amplitude of the effect scaling as $(R_P/a)^2$ \citep{winn:2010}. For example, a 1.5\,$R_{\oplus}$ Super-Earth with a geometric albedo of $A_{g} = 0.5$ orbiting a Sun-like star at 0.05\,AU would have a reflection amplitude below 1\,ppm, making it exceedingly challenging for \kepler\ which has a median sensitivity of ${\sim}40$\,parts per million (ppm) for a 12$^{th}$ magnitude star \citep{christiansen:2013}. Consequently, meaningful constraints on individual albedos are simply unobtainable for almost all Terran and many Neptunian exoplanets found by \kepler.

As noted earlier, a major increase in the SNR is achievable by using phase curves rather than occultations, by essentially increasing the volume of data \citep{kipping:2011}. Similarly, we propose in this work that a further increase in SNR can be achieved by stacking different planets (of similar type) together to create an ensemble phase curve. Although information about the individual planets is lost, this method enables a measurement of the average albedo for a collection of previously unobtainable small worlds.

This stacking approach has been used in numerous other examples recently, such as searching for exotrojans \citep{hippke:2015} and exomoons \citep{teachey:2017}, but the most closely related example comes from 
\citet{sheets:2014} who stack occultations. Stacking occultations is attractive since the effect is intrinsically localized and morphologically sharp, meaning simple local polynomial detrending can be used to correct for trends due to the instrument and stellar activity. However, as highlighted in the case of TrES-2b, occultations have considerably weaker SNR to planetary photons than the full phase curve \citep{kipping:2011}. Nevertheless, \citet{sheets:2014} appear to be first to appreciate the stacking opportunity and in that seminal paper measured an occultation depth of $(3.8\pm1.1)$\,ppm for an ensemble of 31 sub-Saturn ($R<6$\,$R_{\oplus}$) \kepler\ planetary candidates. If the phase curve had zero contribution from thermal emission and was solely due to reflected light, the authors estimate this occultation depth corresponds to an average geometric albedo of $A_g=(0.22\pm0.06)$. In a follow-up study performed by the same authors, similar methods were applied to a larger ensemble, where they measured lower geometric albedos on the order of 0.1 \citep{sheets:2017}.

In this study, we aim to maximize our sensitivity by stacking phase curves to measure the representative albedos of an ensemble of 50 Terran planets and an ensemble of 115 Neptunian planets -- the largest phase curve ensemble considered to-date. In Section \ref{sec:data}, we describe our detrending and stacking methods, and define our target and ensemble criteria. In Section \ref{sec:model}, we provide the expressions used to construct the phase curve models, and in Section \ref{sec:regression} we describe our regression methods and model selection. In Section \ref{sec:results} we state our results for the two ensembles, which are then discussed in Section \ref{sec:discussion}.

%% file: data.tex
\subsection{Frequently used detrending methods}

In this work we seek to measure the phase curve signal of a large ensemble of
\kepler\ exoplanets. Unlike when stacking the transits of many planets (e.g.
\citealt{teachey:2017}) or the occultations (e.g.
\citealt{sheets:2014}), the signal of interest lasts for days and not just
hours when considering full phase curves. Over timescales of days, significant variations are observed in \kepler\
time series, for example due to focus drift and intrinsic stellar activity, and
these trends require removal before the phase curves of each planet can be
co-added. The multi-day nature of the phase curve signal imposes the
requirement of far greater long-term stability in the final data product than
that typically needed when studying transits or occultations. As a result,
many conventional detrending approaches are not well-suited for the task at hand.

For example, polynomial detrending would be inappropriate since even on 
the timescale of a few hours polynomial orders up to $4^{\mathrm{th}}$ order
are often necessary \citep{sandford:2017} for adequate detrending.
Accordingly, for multi-day timescales, very high order polynomials would be
needed, which become increasingly unstable. One solution to this is to use a
moving polynomial kernel (e.g. \citealt{armstrong:2016}), although assuming a
strict polynomial-shape for the variations does not have a clear motivation
besides from mathematical convenience. Further, the order of the polynomial
function must be selected, typically using metrics such as the Bayesian
Information Criterion as a proxy for the full marginal likelihood. A problem
with this is that often competing orders will score similar marginal
likelihoods, meaning that formal Bayesian model averaging across an infinite
sum of polynomial orders is necessary.

Another popular approach is cosine filtering, which was first developed
specifically for phase curves analysis \citep{mazeh:2010}, although it has
since found significant value in transit analysis too (e.g. \citealt{HEK2}). By
adopting a Fourier-based approach, quasi-periodic signals should be expected to
be more accurately described than polynomials which distort the power
spectrum in unpredictable ways. However, since the cosine filtering comprises
of a linear sum of cosine terms of ever higher orders, it too becomes unstable
at high orders often necessary when studying trends over many days \citep{HEK2}.
Similarly, the algorithm typically detrends using a linear sum of harmonic
functions where the number of harmonics is fixed. As with polynomial
detrending, this is somewhat unsatisfactory since increasing or decreasing the
number of harmonics should lead to closely competing maximum likelihoods,
meaning that formal Bayesian model averaging should be invoked.

To overcome these issues, non-parametric detrending methods are
an attractive alternative. An obvious candidate would be median filtering,
where one computes a moving median function at a specific kernel bandwidth over
the entire time series and divides out the time series by this function. One
popular choice for the bandwidth is of order $\sim\mathcal{O}[10^1]$
consecutive points (e.g. \citealt{jenkins:2015}), but this would not be
appropriate here. At such a small size, the method would remove not just the
instrumental and stellar variations, but also the variations due to the phase
curve itself. Rather than use a bandwidth based on consecutive points, it is
often preferable to use bandwidths with fixed temporal windows to account for
potentially sparse data arrays. In this vein, a common choice is a longer
bandwidth arbitrarily fixed to some value of order of days (e.g.
\citealt{rowe:2015}), but such an approach is not designed or indeed intended
to preserve phase curve functions.

\subsection{Non-parameteric detrending with \phasma}

To address the problems discussed above, we designed a new algorithm to detrend exoplanet phase curves that is simple yet powerful, which we refer to as \phasma. Although similar non-parametric methods have been developed for extracting transit signals \citep{samsing:2015}, \phasma\ is instead optimized for phase curve recovery. The philosophy of \phasma\ is that a planetary phase has a highly predictable spectral response function if the planet's orbital period is
a-priori known, as is the case for transiting systems. Specifically, the
phase curve will have strong power at $\omega=2\pi n/P$, where $n$ is the
list of natural numbers to account for harmonics of the phase curve.
This information could be used to design a filter that near-perfectly
removes the planetary phase curve component of the original time series,
$F(t)$, leaving behind a pure nuisance signal, $G(t)$. Finally, the 
nuisance signal can be then removed from the original time series to
reconstruct the planetary phase curve, followed by phase folding and phase
binning to enhance the final data product.

\begin{figure}
\begin{center}
\includegraphics[width=8.4cm,angle=0,clip=true]{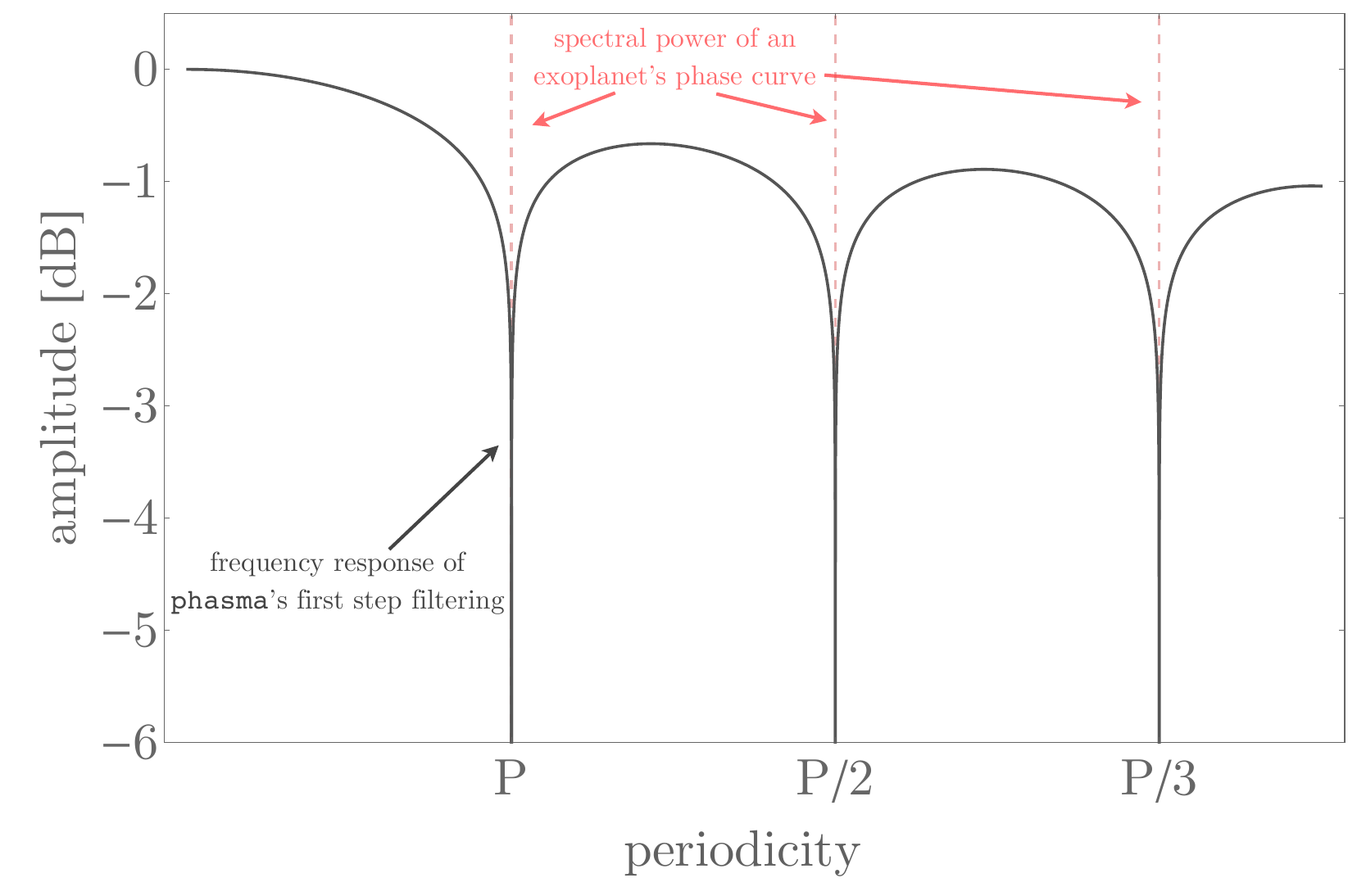}
\caption{
Spectral response function of applying a moving average filter with
bandwidth $P$ to a time series, $F(t)$, representing step 1 of \phasma.
The extreme attenuation at each harmonic removes any flux contributions
from the planetary phase curve, $F_P(t)$, leaving behind a pure nuisance
signal $G(t)$.
}
\label{fig:phasmaresponse}
\end{center}
\end{figure}

Although many band-stop filters could be considered to construct $G(t)$,
an attractive option is the moving average. A moving average of bandwidth
$P$ has excellent attenuation at $\omega=2\pi /P$ and all higher harmonics
(see Figure~\ref{fig:phasmaresponse}). Moreover, the filter is computationally
cheap and conceptually simple. In practice, we elect to use a moving median
filter instead of the mean, since it has the same spectral response properties
but operates as a more robust estimator in the presence of outliers.

To illustrate \phasma, consider taking the first data point of a photometric
time series and then moving out to one orbital period later in time. If the
data were solely due to a phase curve, then this segment of data would
encompass an entire phase curve and thus have a median equal to unity (for a
normalized curve). If the data were due to a phase curve plus some nuisance
signal, then similarly the phase curve contribution averages out and the median
point is simply equal to the median level one would obtain if the signal were
due to the nuisance signal alone. In this way, by moving along point-by-point
with a kernel bandwidth of $P$, we trace out the nuisance signal exclusively.

\phasma\ can be understood as comprising of the following three-step process

\begin{enumerate}
\item Construct a nuisance signal template using coherent median filtering
\item Remove the nuisance signal from the original time series
\item Phase-fold the residual signal to attenuate any non-coherent power
\end{enumerate}

To formally prove the principle behind \phasma, we provide here a mathematical
description of the algorithm by defining the observed flux to be a linear sum
of star's flux, $F_{\star}(t)$, and the planet's flux $F_P(t)$. We explicitly
assume that $F_P(t)$ is a real-valued, strictly periodic function (although it
need not be sinusoidal) and that the waveform of this function does not evolve
in time (e.g. the amplitude does not vary in time), such that it exhibits
translation symmetry:

\begin{align}
F_P(t + n P) &= F_P(t)\,\,\,\forall\,\,\,n \in \mathbb{Z}
\end{align}

assuming the above means that the spectral response function of $F_P$ exhibits
line spectra located at $\omega=2\pi n/P$, where $n$ is a natural number, as
depicted in Figure~\ref{fig:phasmaresponse}.
In contrast, $F_{\star}(t)$ is not assumed to follow any particular functional
form (although it is assumed to be real-valued), since indeed real stars can
produce highly complex and intricate light curves. As mentioned earlier, in
practice, the \phasma\ algorithm uses a moving median, but we consider here
using a moving mean for mathematical convenience, and explain the use of
medians shortly. We further consider the time series to be well-approximated
as being homoscedastic. \phasma\ works by first constructing a nuisance signal
time series, $G(t)$, which may be expressed as

\begin{align}\label{eq:2}
G(t) =& \frac{ \int_{t=t'-P/2}^{t'+P/2} \big[ F_P(t) + F_{\star}(t) \big] \,\mathrm{d}t }{ \int_{t=t'-P/2}^{t'+P/2} \mathrm{d}t },\nonumber\\
\qquad=& \frac{ \int_{t=t'-P/2}^{t'+P/2} F_P(t) \,\mathrm{d}t }{ \int_{t=t'-P/2}^{t'+P/2} \mathrm{d}t } + \frac{ \int_{t=t'-P/2}^{t'+P/2} F_{\star}(t) \,\mathrm{d}t }{ \int_{t=t'-P/2}^{t'+P/2} \mathrm{d}t }
\end{align}

which may be re-written more compactly as

\begin{align}
G(t) &= \bar{F_P} + [F_{\star} \ast \Pi](t).
\label{eqn:Goft}
\end{align}

In reality Equation \ref{eq:2} is performed as a sum of discrete observables, however we argue that the time sampling is much finer than the baseline and therefore any resulting discretization error is negligible. In Equation \ref{eqn:Goft}, $\bar{F_P}$ is time-invariant, since marginalizing any
periodic function over its period returns a constant, which is equal
to the mean planetary flux in this case. In contrast, the
second term preserves time variability, where we have exploited the
fact that a moving average is equal to the convolution of a gate
function, $\Pi(t)$, with $F_{\star}(t)$,

where

\begin{equation}
\Pi(t) =
\begin{cases}
0  & \text{if } |t| > \tfrac{P}{2} ,\\
1 & \text{if } |t| \leq \tfrac{P}{2}.
\end{cases}
\end{equation}

\begin{figure*}
\begin{center}
\includegraphics[width=17.0cm,angle=0,clip=true]{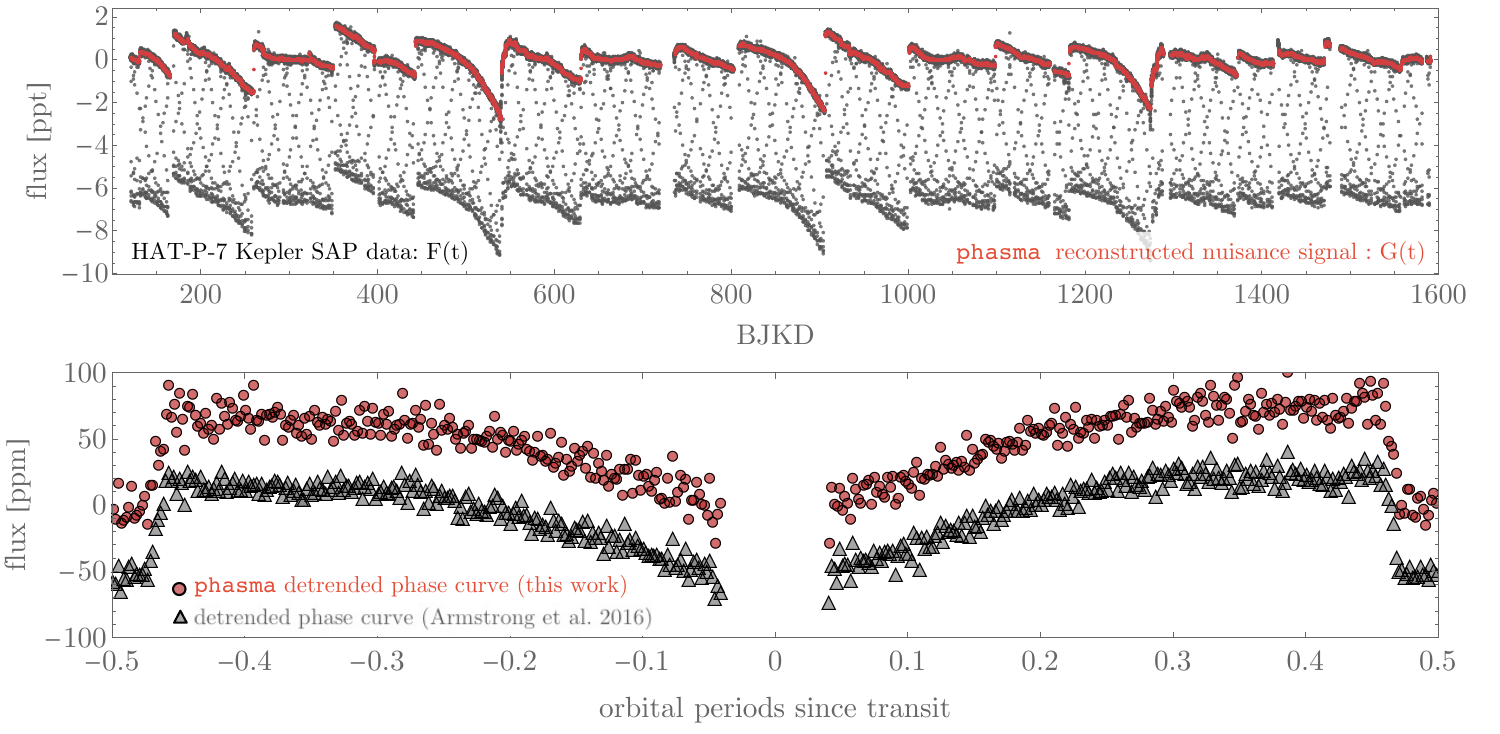}
\caption{
Top panel: SAP light curve of HAT-P-7, transits are clearly visible.
We overlay the moving median function (red) adopting a kernel width
of $P$. Lower panel: Resulting phase curve from our detrending (red cirlces, offset by +50 ppm) compared to the phase curve of \citet{armstrong:2016} (grey triangles) for
the same system but derived using an independent method.
}
\label{fig:hatp7}
\end{center}
\end{figure*}

It is instructive to consider the effect of the kernel on $F_{\star}(t)$ in the
frequency-domain, which can be derived taking the Fourier transform (which
we denote as the operator $\mathcal{F}$) of $F_{\star} \ast \Pi$. Via the
convolution theorem, $\mathcal{F}[F_{\star} \ast \Pi] = 
\mathcal{F}[F_{\star}] \cdot \mathcal{F}[\Pi]$, and it is easy to show that

\begin{align}
\mathcal{F}[\Pi(t)](\omega) &= \frac{P\,\sinc(P\omega/2)}{\sqrt{2\pi}}.
\end{align}

Since the sinc function's amplitude decreases as $\omega^{-1}$, high
frequencies are attenuated, whereas low-frequencies pass through.
Accordingly, let us write that

\begin{align}
F_{\star}(t) &= F_{\star,\mathrm{low}}(t) + \Delta F_{\star}(t)
\end{align}

where the ``low'' subscript denotes the low frequency range (i.e. a periodicity which is greater than the period of the planet) and $ \Delta F_{\star}$
represents the residuals between the original function and its low-pass
filtered component. Since $F_{\star}(t)$ is a positive, real-valued function
at all times (flux cannot be negative), then the moving average must also
be positive and real-valued. By definition, the residuals of the original
function around the moving average must therefore be real-valued but
approximately equally mixed between positive and negative values. This
point will become important when \phasma\ performs a phase-folding operation
later. Returning to Equation~\ref{eqn:Goft}, we may now write that

\begin{align}
G(t) &= \bar{F_P} + F_{\star,\mathrm{low}}(t).
\end{align}

Having constructed and defined $G(t)$, we now proceed to step 2 of \phasma\
and remove $G(t)$ from the original time series. The target function for the
final phase curve function, $\tilde{F}(t)$, is defined in this work as being
the planetary flux divided by the mean stellar flux, $F_P/F_{\star}$,
in the case of perfect detrending. To achieve this, we can write

\begin{align}
\tilde{F}(t) =& \frac{F(t) - G(t)}{G(t)},
\label{eqn:norm}
\end{align}

which is equivalent to taking the original time series, dividing it by
$G(t)$, and then subtracting unity (i.e. normalization + offsetting).
Expanding out these functions using our earlier results, we have

\begin{align}
\tilde{F}(t) =& \frac{ \big[ F_P(t) + F_{\star}(t) \big] - \big[ \bar{F_P} + F_{\star,\mathrm{low}}(t) \big] }{  \big[ \bar{F_P} + F_{\star,\mathrm{low}}(t) \big] },\nonumber\\
\qquad\simeq& \Big(\frac{ F_P'(t) }{ F_{\star,\mathrm{low}}(t) }\Big) + \Big(\frac{ \Delta F_{\star}(t) }{  F_{\star,\mathrm{low}}(t) }\Big),
\end{align}

where on the second line we have defined $F_P'(t) = F_P(t) - \bar{F_P}$
(ultimately a constant offset does not affect our inference of the
phase curve shape) and in the denominator used the $F_{\star,\mathrm{low}}(t)
\gg \bar{F_P}$.

The final step is to take $\tilde{F}(t)$ and phase-fold upon the orbital
period, $P$. Since $F_P'(t)$ is periodic in $P$, and we fold exactly on $P$,
then the signal is completely undisturbed by this process. In contrast, the
$\Delta F_{\star}(t)$ function is, in general, incoherent in $P$ and is
approximately evenly distributed between positive and negative values, by
virtue of its construction (see earlier discussion). Accordingly, randomly
sampling $\Delta F_{\star}(t)$ and taking the mean will converge towards
zero as we increase the number of samples, broadly following a $N^{-1/2}$
scaling. In other words, $\Delta F_{\star}(t)$ averages out on the phase
fold.

We can get a quantitative handle on this attenuation by approximating
$\Delta F_{\star}(t)$ as a high frequency sinusoidal wave, for which the
standard deviation of the signal is $\simeq |\Delta F_{\star}|/\sqrt{2}$,
where $|\Delta F_{\star}|$ is the sinusoidal amplitude. After conducting
upon $B/P$ folds (where $B$ is the baseline of the time series), this
is reduced to  

\begin{align}
\tilde{F}_{\mathrm{fold}}(\phi) =& \Big(\frac{ F_P'(\phi) }{ \bar{F}_{\star,\mathrm{low}} }\Big) + \mathcal{O}\Big[ \frac{\sqrt{P}}{\sqrt{2 B}} \frac{ |\Delta F_{\star}| }{ \bar{F}_{\star,\mathrm{low}} }\Big],
\end{align}

where $\phi$ is the orbital phase of the planet.
Further attenuation of the residual stellar noise is then possible using
phase-bins, since again the planetary signal remains coherent in the
folded domain but the stellar component should not. Despite this,
there remains a possibility for additional noise contamination by the
star and thus we later show how cases with excess noise can be identified
via the use of control system tests (see Section~\ref{sub:control}).

We highlight that in what has been represented thus far, the instrument
response function has been ignored, but it is easy to see that it does not
affect \phasma\ so long as the instrument is dominated by low-frequency
power. This can be seen from Equation~\ref{eqn:norm}, since an instrumental
function acts as a product to $F(t)$ and thus also $G(t)$ and therefore
is cancelled out in that equation, so long as the instrument function is
largely unaffected by the convolution kernel. Quarter stitching is one
way to introduce high frequency power, since step functions between each
quarter would have high power in the frequency domain. For this reason,
this work applies \phasma\ on a quarter-by-quarter basis.

Because the kernel bandwidth is $P$, we cannot reconstruct the nuisance signal in the first and last $P/2$ segments of the time
series. Therefore we require that the baseline $B \gg P$ so that the fraction of unreconstructed data is
relatively small, and thus the method remains both robust and sensitive.

Finally, rather than using a moving average, \phasma\ is run with a moving
median as a more robust estimator for time series featuring outliers and
flares.

\subsection{Suitable \phasma\ targets}

A basic requirement for \phasma\ is a precise measurement of each planet's
orbital period. Since the planets considered in this work are transiting, then
the period is indeed precisely known in all cases. 

Each \kepler\ quarter is offset slightly from the surrounding quarters
due to the rotation of the spacecraft causing the stars to appear on different
CCDs of slightly different sensitivities. An attempt at correcting for this
is made in the PDC-MAP data product \citep{smith:2012,stumpe:2012} although
the associated uncertainty in that correction is unclear. For simplicity, we
detrend each \kepler\ quarter independently, which means $B \sim 90$\,days. In
order to satisfy $B \gg P$ then, our targets are selected
such that $P<10$\,days.

In addition to this filter, we also eliminated systems with more than one
planet, where the phase curves would co-add and negatively affect our
detrending approach. Next, we only considered planets with a NASA
Exoplanet Archive (\href{https://exoplanetarchive.ipac.caltech.edu}{NEA};
\citealt{akeson:2013}) disposition of ``CONFIRMED''. 

Finally, we removed
stars with logarithmic surface gravities less than $4$. The log(g) cut is motivated to remove stars with a higher false-positive rate \citep{sliski:2014} as well as stars with increased activity signals, which have the potential to negatively influence our results. In doing so we attenuate the contamination from these objects and declare a well-defined cut to make our work reproducible. 

Applying these cuts on
September 29th 2016 gave us 477 Kepler Objects of Interest (KOIs) which
we used in what follows.

\subsection{Applied example of \phasma: HAT-P-7b}

As an example of our detrending method, we showcase its performance for
a well-known system, HAT-P-7b, which exhibits strong phase curve variations
\citep{borucki:2009}. The simple aperture photometry (SAP) for this star
is shown in Figure~\ref{fig:hatp7}, where we have normalized each quarter
by the median flux. Setting $P=2.204735417$\,days as reported on
\href{https://exoplanetarchive.ipac.caltech.edu}{NEA}, we derive the
median function, shown in red in the top panel of Figure~\ref{fig:hatp7}, and then divide
the original data through this function. End member data (within $P/2$ of
the quarter stitch points) are not reconstructed. In this example and the real
analyses shown later, only the long-cadence data is used or necessary.

The data are then phase folded, centered upon the time of transit minimum,
and the data binned to 500 evenly spaced phase points. Outliers exceeding
5\,$\sigma$ from each phase window are rejected. The final phase curve
is shown in the bottom panel of Figure~\ref{fig:hatp7} which we compare to that
presented in the independent analysis of \citet{armstrong:2016}. Although
\citet{armstrong:2016} used a moving polynomial based detrending, the
two methods provide equivalent results and give us confidence that our
algorithm performs as expected. Unlike \citet{armstrong:2016}, there is
no need with our method to experiment with different polynomial orders
or kernel sizes, and the detrending options are fixed and objective enabling
a fast and homogeneous detrending of hundreds of KOIs.

\subsection{Choosing a binning scheme}

In order to homogenize the different phase curves of each KOI, we decided to
bin the data into 500 evenly-spaced phase bins. We then removed all phase
points occuring within 5\% of the orbital period centered on the transit to ensure
transits do not contaminate our sample, leaving 450 points.

For a given KOI,
the point-to-point formal uncertainty estimates do not vary greatly throughout
the time series and thus is approximately homoscedastic. We attempted four
different binning schemes to create our binned phase curves for each
individual KOI: a) simple mean b) weighted mean c) simple median d) weighted
median. The weighted-options account for the modest degree of formal
heteroscedasticity.

\begin{figure*}
\begin{center}
\includegraphics[width=17.0cm,angle=0,clip=true]{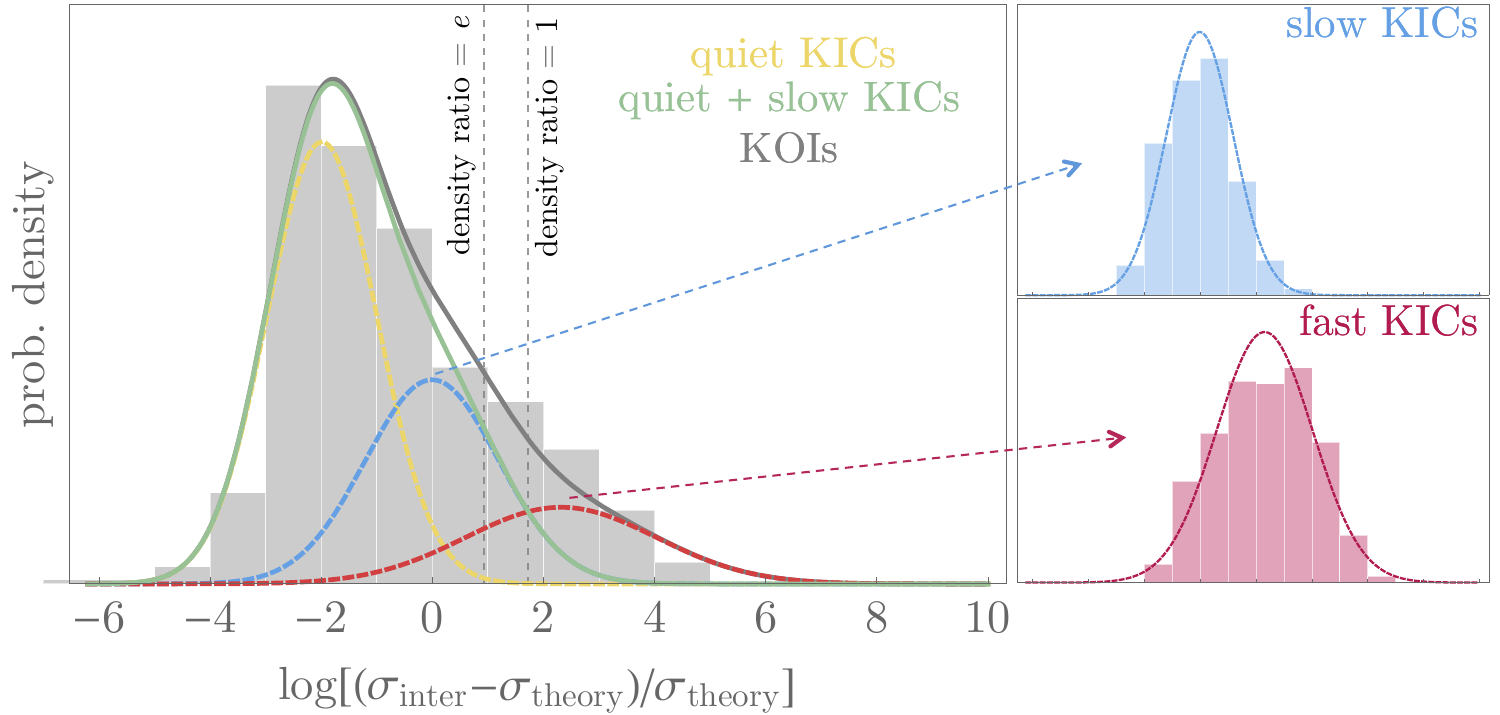}
\caption{
Right panels: Histograms showing the distribution of standard deviations
obtained (normalized by a theoretical benchmark) for the slow rotating
and fast rotating control sample. Left panel: Same as right except for
the 477 KOIs considered in this work. The distribution is well described
by a mixture model of the fast KICs, slow KICs and an additional quieter
subsample. The density ratio of $e$ between the quiet+slow and fast
populations demarks our cut-off for acceptable noise properties.
}
\label{fig:fastslow}
\end{center}
\end{figure*}

We binned all 477 KOIs in our sample with all four methods and then measured
the median absolute deviation of the final binned light curve from each. We
ranked the methods from best to worst for each KOI and then compiled the list
of ranks for all KOIs. We find that the weighted mean is the method which ranks the highest, winning 477 of the 477 trials. Accordingly, we
elected to adopt the weighted mean binning scheme in what follows.

\subsection{Applying to control systems}
\label{sub:control}

After applying our detrending algorithm to all 477 KOIs, we required some
way to assess whether the resulting phase curves were of acceptable quality.
This required a point of comparison and thus we elected to run our code
on two other samples of 477 targets, where no KOIs are presently known to
reside to serve as control systems. The first group is dubbed as ``slow KICs'',
chosen since their rotation periods are slow and thus should be expected to
display lower stellar activity than usual. The second group is dubbed the
``fast KICs'' for the opposite reason, since they have fast rotation and thus
should have significant stellar activity.

The slow sample was defined such that rotation period was slower than 20\,days,
the effective temperature of the star was between 3000\,K and 7000\,K and
the log surface gravity was greater than 4.0. The fast sample used the same
cuts except the rotation period was required to be faster than 5\,days. In both
cases, we used the \citet{mcquillan:2014} rotation periods to perform these
cuts and ensured no KOIs existed at the time of writing.

For the fast sample, we detrended the time series assuming a fictional orbital
period randomly drawn from between 1 and 5\,days. This represents a worst-case
scenario for our method since these fictional orbital periods are nearly in phase with the intrinsic stellar variations. In the case of the slow sample,
we draw a random orbital period from the same distribution, but here the
intrinsic variations are expected to be much slower than the kernel width,
which should lead to a more precise detrending.

After detrending, we define the uncertainty on the phase curve points in two
ways. First, we define $\sigma_{\mathrm{theory}}$ as being the uncertainty
computed from a simple weighted mean calculation, thus equal to

\begin{align}
\sigma_{\mathrm{theory}} \equiv \sqrt{ \sum_{i=1}^n \frac{1}{\sigma_i^{-2}} },
\label{eqn:sigmatheory}
\end{align}

where $\sigma_i$ are the SAP reported photometric errors and the sum is
performed over a specific binning window. We find, as expected, for any
given KOI that the 450 binned $\sigma_{\mathrm{theory}}$ are very similar
and thus simply adopt their median value as a fixed point estimate for each
unique KOI in what follows.

In addition to $\sigma_{\mathrm{theory}}$, which represents a best-case
scenario uncertainty, we also empirically measure the uncertainty in the
final phase curve by computing the standard deviation. Since the phase
curve of each KOI can exhibit significant variations due to the intrinsic
phase curve signal, rather than noise, it is necessary to first perform
a first-order removal of any such signal. To accomplish this, we use
a simple linear least squares fit of composite sinusoidal model for the
ellipsoidal variations, Doppler beaming and reflection+thermal component
following Equation~6 of \citet{mazeh:2010}. We then define $\sigma_{\mathrm{inter}}$ as the the standard deviation of
the residuals.

An idealized system would have $\sigma_{\mathrm{inter}} \simeq
\sigma_{\mathrm{theory}}$, but persistent nuisance variations unaccounted for
by our detrending are generally expected to inflate $\sigma_{\mathrm{inter}}$
above $\sigma_{\mathrm{theory}}$. In this way, we identify the ratio
$(\sigma_{\mathrm{inter}}/\sigma_{\mathrm{theory}})$ as a key metric for
assessing the quality of our detrending. Since in general we expected the best
case to be unity, we subtract one from this ratio and take the log to more
clearly inspect the diversity of ratios found. Repeating for our three samples
(the KOIs, slow KICs and fast KICs), we histogram the resulting ratio proxies
in Figure~\ref{fig:fastslow}.

As can be seen in Figure~\ref{fig:fastslow}, the fast KIC sample exhibits a
larger typical ratio than of the slow KICs, as should be expected, since our
detrending algorithm is most likely to struggle in the fast regime.

Remarkably,
the KOI sample displays an even lower ratio than that of the slow KIC control
sample, implying that KOI host stars are less active than even the slow
rotating \kepler\ stars. One possible explanation for this could be that a population of non-transiting planets are contaminating the slow KIC sample, thus generating a higher uncertainty ratio, however we find  this explanation improbable. The number of transiting exoplanets that produce independently detectable phase curve amplitudes has been relatively low, and therefore we would expect to see a similar paucity of non-transiting planets with non-negligible phase curves in our slow KIC sample. Additionally, the phase curve amplitude is at its maximum when the orbit is along the line of sight and would therefore decrease for non-transiting exoplanets, further lowering the possibility of being responsible for the larger uncertainty ratio in the slow KIC sample. Instead, this is most likely an artifact of selection bias,
since transit signals are intrinsically easier to detect around quieter stars.

We find that normal distributions well-describe both the slow and fast KIC
histograms in Figure~\ref{fig:fastslow} and thus one can treat the KOI sample
as being a mixture model of these two components plus some additional even
quieter component. Assuming this additional component is also a normal
distribution, we use maximum likelihood regression to find that the KOI
sample is well-described by a mixture model of 16.2\% fast KICs, 30.0\% slow
KICs and 53.7\% an additional quiet sample component.

In general, we deem it satisfactory if a target star can be said to belong
to either the quiet or slow KIC sample, since our detrending algorithm is
designed with such cases in mind. Accordingly, using our mixture model,
we demark that for all $(\sigma_{\mathrm{inter}}/\sigma_{\mathrm{theory}})
< 6.6$, it is more likely a sample would belong to the slow+quiet sample
than the fast sample. At the critical ratio of 6.6, the probability density
ratio between the two populations is unity, and thus for the sake of
conservatism we push this back until the probability density ratio equals $e$,
which occurs at $(\sigma_{\mathrm{inter}}/\sigma_{\mathrm{theory}}) = 3.6$.

\begin{figure}
\begin{center}
\includegraphics[width=8.4cm,angle=0,clip=true]{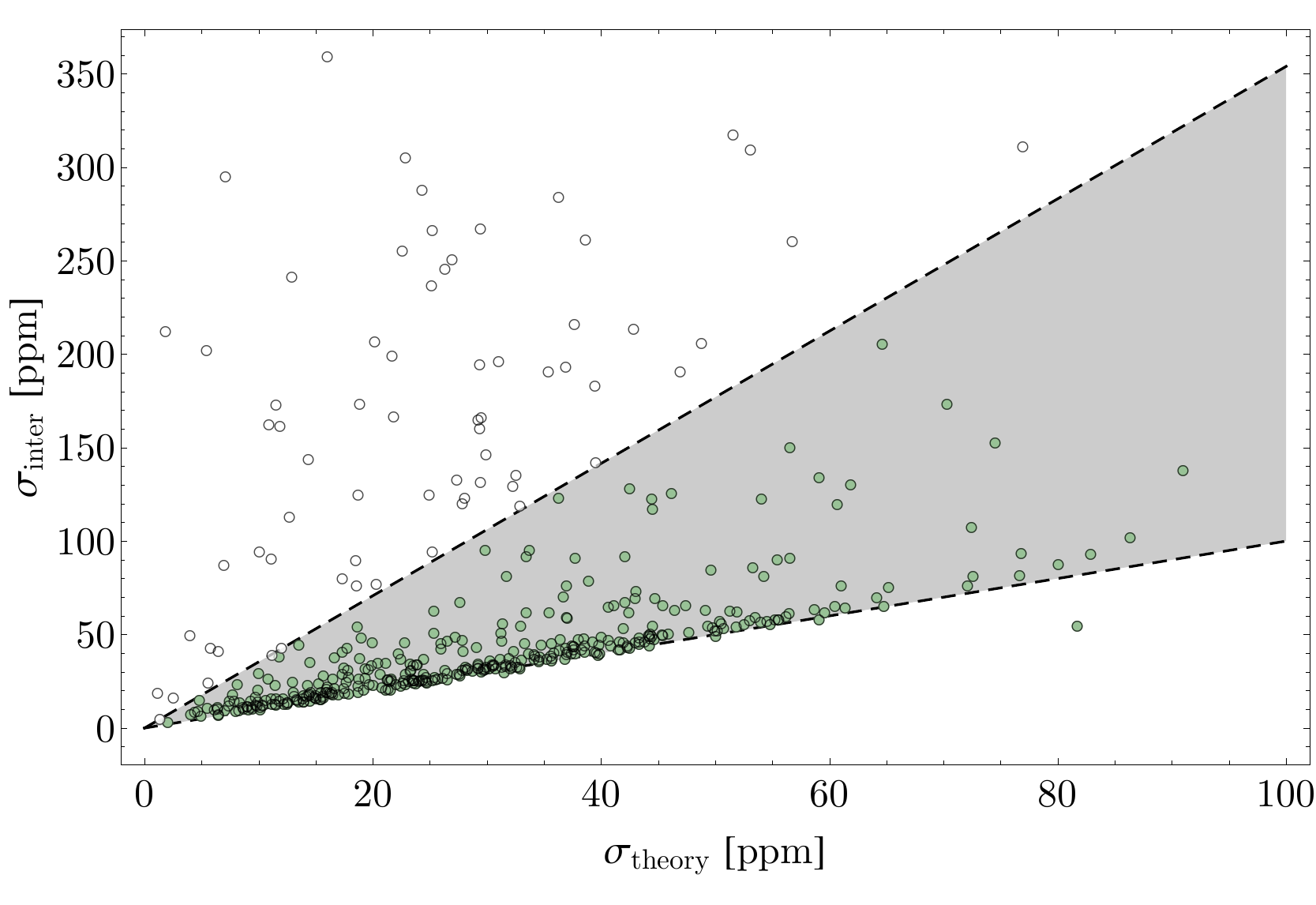}
\caption{
Scatter plot of the observed standard deviation versus the theoretical
prediction with perfect detrending for the 477 KOIs considered in this work.
Objects south of a ratio of 3.6 (upper dashed line) are considered to be of
acceptable quality in what follows.
}
\label{fig:noisecut}
\end{center}
\end{figure}

In Figure~\ref{fig:noisecut} we show a scatter plot of the two noise estimates
for the 477 KOIs and draw a line demarking the cut-off ratio of 3.6. This cut
reduces our sample from 477 KOIs to 378, removing 99 targets.

\subsection{Stacking different KOIs}
\label{sub:stacking}

Each binned phase curve has a slight but somewhat arbitrary offset that
means simple stacking of the curves present noticeably poor coherence. It is
therefore necessary to define an offset term for each KOI which is subtracted
prior to the final stacking.

Consider the $j^{\mathrm{th}}$ binned phase point of the final stacked phase
curve of all KOIs, where $j=1,2,...,n_{\mathrm{bins}}-1,n_{\mathrm{bins}}$
(where in our case $n_{\mathrm{bins}}=450$). Within that bin, there are
$n_{\mathrm{KOI}}$ data points, one from each KOI, contributing to the binned
point. Ideally they would closely agree with each other and display a small
spread. If the spread were large, this would indicate we have poorly selected
our offset terms. Formally, rather than spread the standard deviation, we
really want to minimize the chi-squared of the $n_{\mathrm{KOI}}$ points away
from the average; $\chi_j^2$. Since we can do this for each and every binned
phase point ($n_{\mathrm{bins}}=450$ in total), the final cost function is the
sum of all of these chi-squared terms i.e. $C = \sum_{j=1}^{n_{\mathrm{bins}}}
(\chi_j)^2$. In principle then, we simply need to minimize $C$
with respect to $n_{\mathrm{KOI}}$ offset parameters, which we may write as

\begin{align}
C(\boldsymbol{\theta}) &= \mathlarger{\mathlarger{\sum}}_{j=1}^{n_{\mathrm{bins}}} \mathlarger{\mathlarger{\sum}}_{i=1}^{n_{\mathrm{KOI}}} \Bigg( \frac{(f_j^i - \theta^i) - \mu_j}{\sigma_j^i} \Bigg)^2,
\end{align}

where we use $i$ superscript to denote the $i^{\mathrm{th}}$ KOI and $j$
subscript to denote the $j^{\mathrm{th}}$ binned phase point, and we
further define $\theta^i$ as the offset term associated with the
$i^{\mathrm{th}}$ KOI, and $\mu_j$ as the average of the $j^{\mathrm{th}}$
binned phase point.

One complication is that within each grand bin point, we want the points to lie
close to the ``average'', $\mu_j$, but we can again choose to define average in
several different ways: a) simple mean b) weighted mean c) simple median d)
weighted median. Between KOIs, the data is certainly heteroscedastic here and
thus one should not expect the unweighted versions to be optimal. Nevertheless,
we tried all four strategies on each planet group considered. We defined the
prefered method as that which leads to the lowest median absolute deviation in
the final grand stacked phase curve. For all planet groups considered, we found
that the weighted mean was indeed the preferred metric, such that

\begin{align}
\mu_j &= \frac{ \sum_{i=1}^{n_{\mathrm{KOI}}} (f_j^i - \theta^i) (\sigma_j^i)^{-2} }{ \sum_{i=1}^{n_{\mathrm{KOI}}} (\sigma_j^i)^{-2} }.
\end{align}

\begin{figure*}
\begin{center}
\includegraphics[width=17.0cm,angle=0,clip=true]{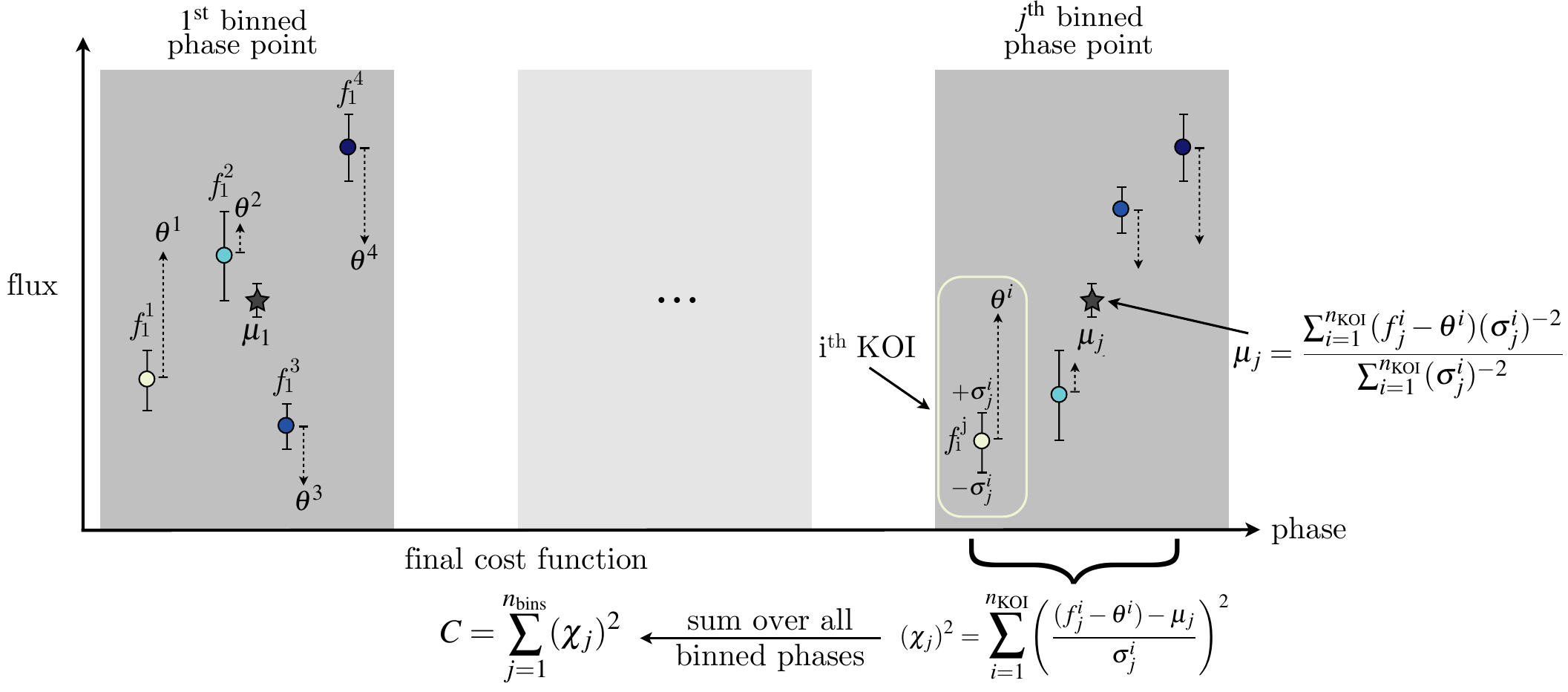}
\caption{
Illustration of the offset optimization scheme used between KOIs.
Each KOI's phase curve is offset by a constant $\Theta^i$ for
the $i^{\mathrm{th}}$ KOI, which is then optimized for by
minimizing the cost function $C$ depicted.
}
\label{fig:offsets}
\end{center}
\end{figure*}

Finally, we point out that for $\sigma_j^i$, we assume that for a particular
given KOI, the errors are homoscedastic (although heteroscedastic across
different KOIs) such that $\sigma_j^i = \sigma^i$ for all $j=1,2,...,
n_{\mathrm{bins}}-1,n_{\mathrm{bins}}$. Specifically, we set $\sigma^i$ to
be equal to the standard deviation of the residuals of each KOI's phase curve
after an initial simple fit (as described earlier in
Section~\ref{sub:control}), such that $\sigma^i$ is equal to
$\sigma_{\mathrm{inter}}$ of the $i^{\mathrm{th}}$ KOI. An illustration of this offset optimization scheme can be seen in Figure~\ref{fig:offsets}.

The uncertainty on the binned points is defined by 1.4826\footnote{The 1.4826 value is a known scaling factor on the median absolute deviation (MAD) of a normal distribution that is needed in order to use the MAD as an estimate of the typical standard deviation \citep{huber:1981}.} multiplied by the median absolute deviation (MAD) of the final binned ensemble curve, after excluding phase points interior to within 5\% of the orbital period centered on the transit. We opt to use the MAD as it is more robust against outliers than the typical standard deviation.

As a brief aside, the optimization of $C$ is non-trivial due to the large
dimensionality and non-linearity of the problem. To solve this, we initially
set the vector of offset terms, $\boldsymbol{\theta}$, to be zero and computed
$C(\boldsymbol{\theta})$. We then replaced the first term, $\theta^1$, with a
variable $x$ and performed a simple downhill 1D minimization of
$C(\boldsymbol{\theta})$ with respect to $x$. We then replaced $\theta^1$
with this optimal value and sequentially repeated for all elements of
$\boldsymbol{\theta}$. After completion, we saved the new $C$ and then repeated
the entire process again. This was done multiple times and we found rapid
convergence of $C$ to machine precision after just a few full iterative rounds.

\subsection{Defining planet samples}
\label{sub:samples}

The objective of this work is to measure the reflection component of the phase
curve for a co-added sample of KOIs. Since phase curves are comprised of
several other effects besides reflection, it is desirable to choose a subset
for which these other effects are expected to have relatively little
contribution. However, the amplitudes of each component are, a-priori, unknown
to us although they can be predicted using empirically calibrated models.

To this end, we forecasted the amplitude of the ellipsoidal variations $A_{\text{ellip}}$, the
beaming effect $A_{\text{beam}}$, the thermal component $A_{\text{thml}}$, and the reflection phase curve $A_{\text{refl}}$ for each
and every KOI. First, we obtained the posterior distributions for each
planet's observed radius and forecasted mass from the study of
\citet{chen:2017b}. The radius posteriors come from combining the
\citet{mathur:2017} stellar posteriors and the \citet{rowe:2015} transit
parameter posteriors together, whereas the forecasted masses come from the
\forecaster\ package \citep{chen:2017a}. We then computed the forecasted
beaming, ellipsoidal, reflection and thermal components. For all calculations,
we assume zero eccentricity for simplicity.

For the beaming amplitude, we use Equation~9 of \citet{mazeh:2010}, such that

\begin{align}
A_{\mathrm{beam}} &= 4 \alpha_{\mathrm{beam}} \tfrac{K}{c},
\end{align}

where $K$ is the forecasted radial velocity amplitude, which can be computed
from the forecasted planetary mass, $c$ is the speed of light in a vacuum and
$\alpha_{\mathrm{beam}}$ is a factor of order unity to account for the finite
bandpass used and the Doppler boosting effect by the spectrum shifting in and
out of said bandpass \citep{loeb:2003}. For this factor, we computed
$\alpha_{\mathrm{beam}}$ using the {\tt IDL} code of {\tt alpha\_beam} by
Brian Jackson (private communication) which depends upon a single input,
the effective temperature of the parent star. We computed
$\alpha_{\mathrm{beam}}$ across a grid from $3000$\,K to $10000$\,K using the
\kepler\ bandpass and found that a fourth-order polynomial - which is
substantially faster to call - provides an excellent approximation, such that

\begin{align}
\alpha_{\mathrm{beam}} \simeq& (7.89) + ( -2.64\times10^{-3} ) T_{\mathrm{eff}}
+ ( 4.33\times10^{-7} ) T_{\mathrm{eff}}^2 \nonumber\\
\qquad& + ( -3.46\times10^{-11} ) T_{\mathrm{eff}}^3
+ ( 1.07\times10^{-15} ) T_{\mathrm{eff}}^4.
\end{align}

For the ellipsoidal variations, we adopt Equation~7 of \citet{mazeh:2010},
which is based upon the approximation of \citet{morris:1993}, such
that

\begin{align}
A_{\mathrm{ellp}} &= \alpha_{\mathrm{ellp}} \frac{M_P}{M_{\star}} \Big(\frac{R_{\star}}{a}\Big)^3,
\end{align}

where $\alpha_{\mathrm{ellp}}$ is a coefficient well-approximated by

\begin{align}
\alpha_{\mathrm{ellp}} &= 0.15 \frac{(15+u)(1+g)}{3-u},
\end{align}

where $u$ is the linear limb darkening coefficient and $g$ is the stellar
gravity darkening coefficient. For these coefficients, we queried the
theoretical tabulation presented in \citet{claret:2011} for the \kepler\
bandpass given a vector of inputs defined by $\{T_{\mathrm{eff}},\log g,
[\mathrm{Fe}/\mathrm{H}]\}$. In order to draw intermediate points not
present in the table, we trained a random forest interpolative algorithm
on the three inputs, enabling us to quickly interpolate
$\alpha_{\mathrm{ellp}}$ for any given choice of inputs.

For the reflection and thermal components, we adopt a fairly conservative
choice for the geometric albedo equal to $A_g = 0.1$. Assuming a Lambertian
sphere, this sets the Bond albedo to $A_b = \tfrac{3}{2} A_g$. We computed
the day- and night-side temperatures of the planet using the prescription of
\citet{cowan:2011b}, such that

\begin{align}
T_{\mathrm{day}} &= T_0 (1-A_b)^{1/4} \big( \tfrac{2}{3} - \tfrac{5}{12} \beta \big)^{1/4},\\
T_{\mathrm{night}} &= T_0 (1-A_b)^{1/4} \big( \tfrac{1}{4} \beta \big)^{1/4},
\end{align}

where $T_0$ is the temperature of the planet at the substellar point
\citep{hansen:2008}, given by

\begin{align}
T_0 \equiv T_{\mathrm{eff}} (a/R_{\star})^{-1/2}.
\end{align}

The $\beta$ term here represents a redistribution factor that we set to
$\beta=\tfrac{1}{3}$, as a low but not unreasonable value in order to
maximize the thermal component of the phase curve. The thermal component was
found by numerically integrating a linearly interpolated high-resolution
tabulation of the \kepler\ bandpass multiplied by the Planck function for the
planetary and stellar components respectively, and then taking the ratio
multiplied by the ratio-of-radii squared. The flux-ratio integration is time
consuming ($\sim0.1$\,seconds per call) and given the large number of calls
needed ($\sim$20 million) we decided to create an initial library of results
from $T_{\mathrm{eff}}=2000\to10000$\,K and $T_{P}=300\to4000$\,K on 30\,K
steps, which we then bicubic-spline interpolated later during the actual
calculations on the posterior samples. Since we use a bicubic grid, we
increased the edges of the grid by $3^2$ grid points either side of our
formal interpolation range to avoid boundary errors.

\begin{figure}
\begin{center}
\includegraphics[width=8.8cm,angle=0,clip=true]{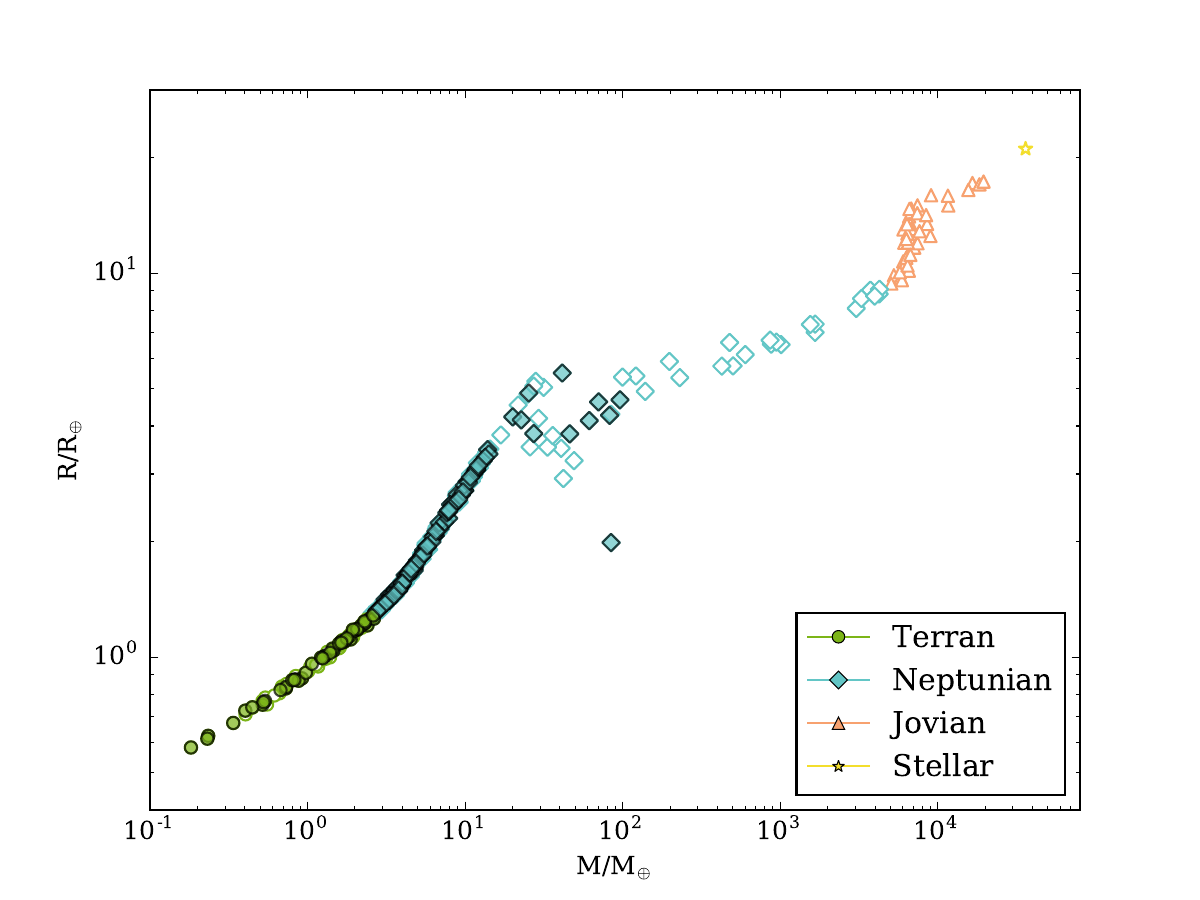}
\caption{
Classifications of each KOI using \forecaster\ taken from \citet{chen:2017b}
for the 477 initial KOIs considered in this work. Classifications shown
represent the modal class probability. Filled points were assigned
as being ``cool'', ``quiet'' and ``light'' (see Section~\ref{sub:samples}
for explanation of these terms), as well as having a $<10$\% chance of
being Jovian or Stellar.
}
\label{fig:classifications}
\end{center}
\end{figure}

Finally, the reflection component is simply computed as $A_{\mathrm{refl}} =
A_g p^2 (a/R_{\star})^{-2}$, where $p^{2}$ is the normalized transit depth. For each KOI, we step through the 40,000 joint
posterior samples from \citet{chen:2017b} and compute the
corresponding set of phase curve amplitudes at each step. For each KOI, we also have
a classification probability (Terran, Neptunian, Jovian or Stellar) based off
the posteriors and \forecaster\ prediction of \citet{chen:2017b}. These
classifications are illustrated in Figure~\ref{fig:classifications} for our
KOIs under consideration.

We first defined a sample of sub-Jovians by using the \forecaster\
classifications for which there is a $\geq 90$\% probability of the KOI being
either Terran or Neptunian. Of these, we then split into Terran and Neptunian
lists if the corresponding class probability exceeded 50\%. For each KOI, we
inspected the posterior samples of the forecasted $A_{\mathrm{refl}}$ and 
$A_{\mathrm{thml}}$ and counted the fraction of samples for which
$A_{\mathrm{refl}}>10^{1/2} A_{\mathrm{thml}}$. If this fraction exceeded 90\%,
we denote the planet as a ``cool'' KOI. Similarly, we counted the fraction of
samples for which $(A_{\mathrm{refl}}+A_{\mathrm{thml}})>10^{1/2} 
\mathrm{max}[A_{\mathrm{beam}},A_{\mathrm{ellp}}]$ and those exceeding 90\%
fractions were labeled as ``light'' KOIs.\footnote{We use $10^{1/2}$ because it is half an order-of-magnitude, and therefore becomes an order-of-magnitude when adding components in quadrature.} The final label we considered were KOIs
for which we attribute the noise properties as most likely belonging to the
quiet- or slow-like KIC samples; objects which for simplicity we dub as
``quiet'' (see Section~\ref{sub:control} for details). Finally, we inspect the
phase curves of each KOI by eye and exclude any which vary obviously from the
overall sample.

We briefly highlight that we make no effort to account for or exclude KOIs
with transit timing variations (TTVs). Short-period planets such as ours rarely
exhibit TTVs and those that do are typically sufficiently low-amplitude to
have negligible effect on the results presented here \citep{mazeh:2013}.

\begin{figure*}
\begin{center}
\includegraphics[width=17.0cm,angle=0,clip=true]{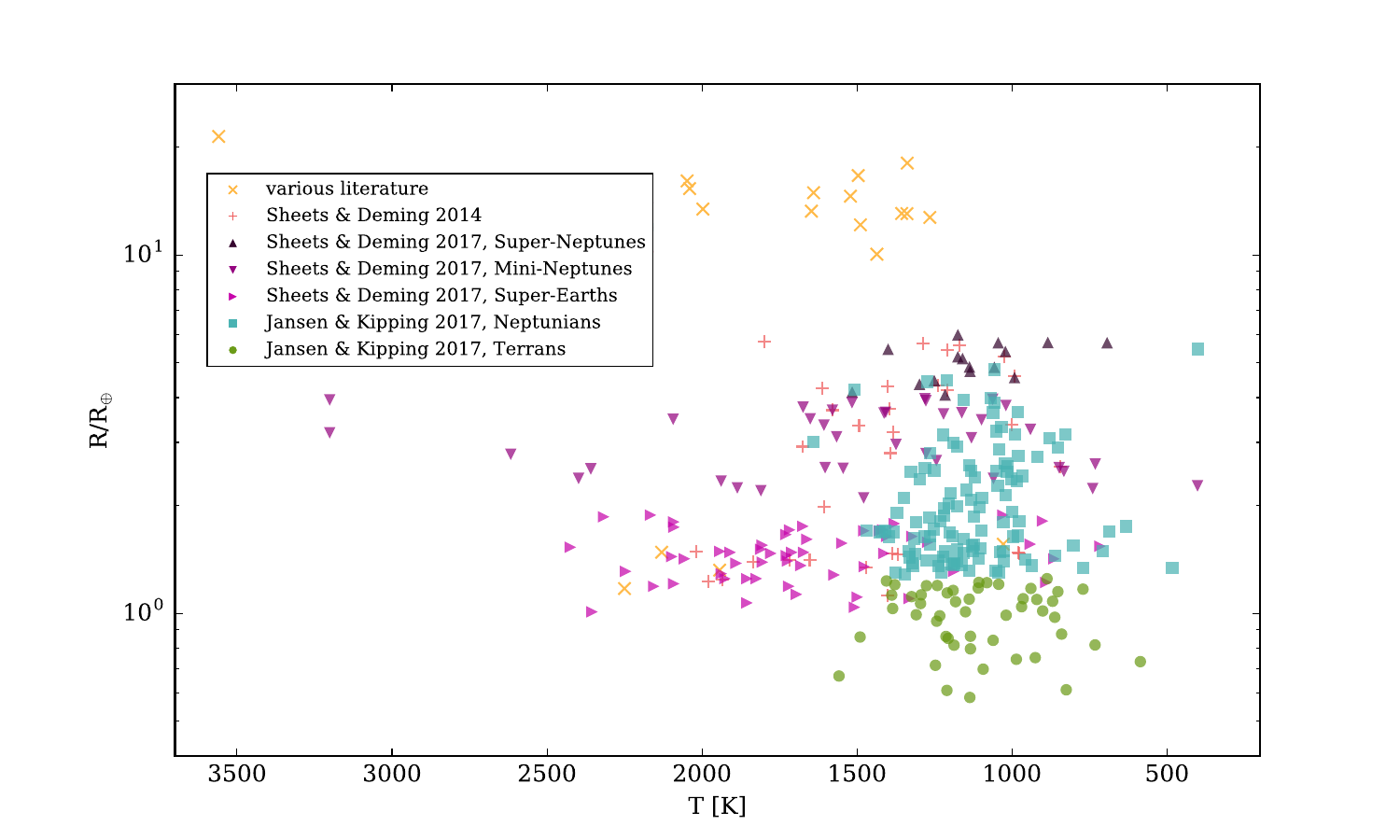}
\caption{
Illustrative comparison of our Terran and Neptunian sample versus that
from \citet{sheets:2014} and \citet{sheets:2017}, as well as other
literature sources \citep{desert:2011,santerne:2011,fortney:2011,batalha:2011,
demory:2011,esteves:2013,sanchis:2013,shporer:2014,deleuil:2014,demory:2014,
gandolfi:2015,angerhausen:2015,armstrong:2016}
}
\label{fig:targets}
\end{center}
\end{figure*}

In total, our filters give us 115 quiet, light, cool Neptunians and 50 quiet,
light, cool Terrans. For comparison with previously studied samples, we plot
the size and equilibrium temperature of our sample in Figure~\ref{fig:targets}.

%% file: model.tex
To measure the representative reflection and thermal quantities of the ensembles, we first generate model phase curves from the sum of a reflection component and a thermal component. The reflection component is assumed to be symmetric for the ensemble, and is proportional to the Bond albedo $A_{B}$ under the Lambertian approximation. The thermal component then depends on the thermal redistribution efficiency factor $\epsilon$ and the greenhouse factor $f$. 

The thermal redistribution efficiency is here defined to be the ratio between the radiative timescale of the planet's photosphere and the difference between the frequencies at which the photosphere rotates about the planet and the surface rotates about its axis. In other words, if the atmospheric mass heated at the substellar point is redistributed about the surface much faster than the heat gets reradiated, the planet would be described as having a large redistribution efficiency $\epsilon$, typically $\epsilon\gg1$. Conversely, a planet with relatively no heat redistribution would be described as having $\epsilon$ = 0. For a planet which has winds moving in a direction opposite of the planetary rotation, $\epsilon$ is defined to be negative. The greenhouse factor $f$ is simply a scaling factor which accounts for any temperature boost due to the presence of greenhouse gases in the atmosphere. 

Our model described in the following section allows us to compute the phase curve of a single exoplanet accounting for thermal emission and reflection. However, as described in Section~\ref{sec:data}, the final data product under analysis is an ensemble of many exoplanets. In what follows, we will assume that each planet within a subset shares the same Bond albedo, $A_B$, thermal redistribution factor, $\epsilon$ and greenhouse factor, $f$. These parameter inferences should be interpreted as measurements of the ``typical'' or ``representative'' values, since in reality there will be an underlying and unknown distribution of these terms.

\subsection{Thermal component}

We express the thermal emission component of the normalized planetary flux as
\begin{equation}
\begin{split}
     F_{T} = &\frac{1}{\pi B_{K,\star}}\left(\frac{R_P}{R_{\star}}\right)^{2}\\
     &\times\int_{-\frac{\pi}{2}}^{\frac{\pi}{2}}\int_{-\frac{\pi}{2}}^{\frac{\pi}{2}}B_{K,P}[T(\alpha, \theta, \phi)]\cos^{2}{\theta}\cos{\phi}d\theta d\phi
\end{split}
\end{equation}
where $B_{K,\star}$ is the Planck function of the host star convolved with the Kepler bandpass, $R_P$ is the radius of the planet, $R_{\star}$ the radius of the star, and $B_{K,p}[T(\alpha,\theta,\phi)]$ is the temperature distribution-dependent blackbody curve of the planet convolved with the Kepler bandpass,
\begin{equation}\label{eq:planck}
\begin{split}
    B_{K,P}[T(\alpha,\theta,\phi)] = &\int_{\lambda}K_{\lambda}\frac{2hc^{2}}{\lambda^{5}}\\
    &\times\left[\exp{\left(\frac{hc}{\lambda k_{B}}\frac{1}{ T(\alpha,\theta,\phi)}\right)} - 1\right]^{-1}d\lambda
\end{split}
\end{equation}
where $K_{\lambda}$ is the Kepler response function, $T(\alpha, \theta, \phi)$ is the phase-dependent temperature distribution across the planet's surface, and where $\alpha$, $\theta$ and $\phi$ represent the planetary phase, latitude and longitude as viewed in the observer's frame of reference, respectively. For our models we have chosen a surface resolution of 15$^{\circ}\times$15$^{\circ}$ in latitude and longitude, where further increasing the resolution changes the thermal amplitude on the order of one-hundredth of a percent.
It should be noted that $\phi$ and $\theta$ are independent of phase, where $\phi \equiv 0$ in the direction of the observer.

We borrow from \cite{hu:2015} to define the phase-dependent temperature distribution $T(\alpha, \theta, \phi)$ to be equal to
\begin{equation}
    T(\alpha, \theta, \phi) = fT_{0}(\theta)\mathcal{P}(\epsilon, \xi)
\end{equation}
where $f$ is the greenhouse boosting factor, $T_{0}$ is the sub-stellar temperature, and $\mathcal{P}$ is the thermal phase function, which for a planet on a circular orbit can be expressed by Equation (10) in \cite{cowan:2011a}:
\begin{equation}\label{eq:therm_P}
    \frac{d\mathcal{P}}{d\xi} = \frac{1}{\epsilon}(\text{max}(\cos{\xi}, 0) - \mathcal{P}^{4})
\end{equation}
where max(cos $\xi$, 0)  = $\frac{1}{2}(\cos{\xi} + |\cos{\xi}|)$, i.e. a cosine function truncated at negative values. We borrow our notation from \cite{hu:2015}, where $\xi$ represents the local planetary longitude defined for all points in phase to be $\xi \equiv \phi - \alpha$. The phase term $\alpha$ ranges from $-\pi$ to $\pi$ and is defined to be zero at the secondary eclipse. For a planet with prograde rotation, $\xi = 0$ at the sub-stellar longitude, $\xi = -\pi/2$ at the dawn terminator, and $\xi = \pi/2$ at the dusk terminator.

Equation (\ref{eq:therm_P}) does not have an analytic solution, so we solve it numerically using \scipy's ODE integrator, where we set the initial conditions equal to the approximated expression for $\mathcal{P}_{\text{dawn}}$ stated in the Appendix of \cite{cowan:2011a},
\begin{equation}
    \mathcal{P}_{\text{dawn}} \approx \left[\pi + (3\pi/\epsilon)^{4/3}\right]^{-1/4}.
\end{equation}

The sub-stellar temperature as a function of planetary latitude $\theta$ is expressed by
\begin{equation}
    T_{0}(\theta) = T_{\star}\left(\frac{R_{\star}}{a}\right)^{1/2}(1 - A_{B})^{1/4}\cos{\theta}^{1/4}
\end{equation}
where $T_{\star}$ is the effective temperature of the host star and $a$ the semi-major axis.

\subsection{Reflection component}

The reflection component of the phase curve is given by
\begin{equation}
    F_{R} = \left(\frac{R_P}{a}\right)^{2}\frac{2}{3}A_{B}\frac{1}{\pi}\left[\sin{|\alpha|} + (\pi - |\alpha|)\cos{|\alpha|}\right]
\end{equation}
where we adopt the Lambertian approximation so that $A_{B} = \frac{3}{2}A_{g}.$ According to \cite{seager:2000b} and \cite{cahoy:2010}, this is a fine approximation under the homogeneously reflecting atmosphere assumption.

%% file: regression.tex
\subsection{Constructing a likelihood}

To infer a subset's atmospheric parameters $A_{B}$, $\epsilon$, and $f$, we require a likelihood function to describe the probability of obtaining the ensemble data given a particular realization of an ensemble model. To create an ensemble model for a particular choice of the atmospheric properties, $F_{\mathrm{tot}}(A_B,\epsilon,f)$, we first generate the phase curves of each individual planet using a set of global atmospheric parameters, denoted by $F_{i,\mathrm{tot}}(A_B,\epsilon,f)$ where $i$ is the planet index. We next take the weighted average of the individual phase curves using the same weighting as that used for the real data stacking (see Section~\ref{sub:stacking}). Finally, this ensemble model is then subtracted from the ensemble data to calculate the residuals, $r_i$. With the residuals in hand, we write a likelihood function by assuming that the ensemble data points are independent and normally distributed, such that

\begin{align}
\log\mathcal{L} &= -\sum_{i=1}^N \frac{1}{2} \log(2\pi) - \sum_{i=1}^N \log \sigma_i
- \frac{1}{2} \sum_{i=1}^N \Big(\frac{r_i}{\sigma_i}\Big)^2.
\end{align}

We make two modifications to our model beyond that described thus far. First, we allow our likelihood function to account for the possibility of underestimated uncertainties. Recall that the measurement uncertainties on each binned point are computed using the median absolute deviation yet we acknowledge the possibility that these may underestimate the true value.  We therefore add a ``jitter'' term, $\sigma_{\mathrm{jitter}}$, in quadrature to the errors, which is itself treated as a free parameter in the model, similar to the prescription described by \citet{teachey:2017}. Second, although our data has been carefully normalized and offset (see Section~\ref{sub:stacking}), we include an offset term to the final ensemble as a free parameter, $\gamma$, which simply allows us to propagate the uncertainty of the offset into the covariances of any resulting posteriors and serve as a final check that the data are indeed normalized correctly.

\subsection{Model Look-Up Tables (LUTs)}

In practice, we found that calling our \python\ implementation of the phase curve model for each planet at each phase point was sufficiently computationally expensive to make conventional Bayesian regression too time consuming. To solve this, we elected to build a pre-computed look-up table (LUT) of ensemble model phase curves across a three-dimensional grid of $A_B$, $\epsilon$ and $f$. For any given choice of these parameters, we can then conduct a tri-linear interpolation of the regular grid to reproduce any choice we wish that falls within the LUT's calibrated range.

The ensemble models take into account the uncertainties in the measurements of stellar radius, mass, effective temperature, and density by sampling from corresponding posterior distributions of these quantities given in \citet{mathur:2017}, generating model phase curves from each sample draw, then implementing the average of these phase curves.

The parameter space from which we generate the models spans a uniform range of $A_{B}$ from [0.0, 1.0] at a resolution of $\Delta A_{B} = 0.01$, $f$ from [1.0, 2.0] at a resolution of $\Delta f = 0.1$, and a log uniform range of $\log\epsilon$ from [-1.7, 1.7] at a resolution of $\Delta\log\epsilon = 0.05$. It is possible for the redistribution factor to extend beyond these limits to infinity, however we consider the change in the thermal component for $|\epsilon| > 50$ to be negligible.

\subsection{Bayesian regression and priors}

We utilize \emcee\ \citep{dfm:2013} to perform an Affine Invariant Markov chain Monte Carlo (MCMC) procedure on the ensemble phase curves. We obtain $2\times10^6$ samples using 100 walkers, burning the first half of the chain for a remaining total of $10^6$ samples. Our priors span a uniform range of $A_{B}$ from [0.0, 1.0], $f$ from [1.0, 2.0], and $\epsilon$ from [-50, 50]. Although we generated the models spanning a log-uniform range of $\epsilon$, sampling from a uniform prior distribution is sufficient for this parameter due to its low impact on the model at $\epsilon \gtrsim$ 10.

\subsection{Savage-Dickey density ratio (SDDR)}
\label{sub:SDDR}

For a model which contains all of the parameters of another plus at least one additional parameter, \citet{dickey:1970} show that the Bayes factor between the nested model and its parent model can be estimated by simply taking the ratio of the probability density of the full model at the null point of the nested model to the prior distribution at that same point, also known as the Savage-Dickey density ratio (SDDR).

For the purpose of completeness, we use the SDDR to compare the results of a fit to a ``black planet'' model (i.e. $A_{B}=0$) to the results of the fit to the full reflection-thermal model. Because the black-planet model is a nested case of the full model, computing the SDDR is an appropriate method of odds comparison for these models. Further, since the black-planet model is separated by just one free parameter from the full model, the density of samples is sufficiently high to accurately resolve the SDDR.

For a probability density $\mathcal{P}$ and uniform prior distribution of the full model $\Pi_{\text{full}}$, we can estimate the Bayes factor with the SDDR to be

\begin{align}
B_{\text{black:full}} &= \frac{\mathcal{P}_{\text{full}}(A_{B} = 0)}{\Pi_{\text{full}}(A_{B} = 0)}.
\end{align}

%% file: results.tex
\subsection{Terran ensemble}

Our model accounts for both the thermal and reflected light component of an
exoplanetary phase curve, and there is a certain degree of trade-off between
the two. This degeneracy means that low signal-to-noise phase curves, as
might be expected for small planets, can have marginalized posterior
distributions for albedo and greenhouse factor which, when taken independently,
appear consistent with zero and thus a null detection. Accordingly, when
considering the basic question as to whether any kind of signal is detected or
not, inspection of marginalized one-dimensional model parameter posteriors is
not the most robust tool. Instead, we argue that it is better to compute the
amplitude of the phase curve directly at each posterior sample and then
construct an a-posteriori amplitude distribution when evaluating detection
significance. Since the phase curve model is non-sinusoidal, a simple amplitude
estimate is not directly available, but we can easily compute the root mean
square (RMS) amplitude of any given phase curve model, which is suitable as
a proxy for signal amplitude.

\begin{figure}
\includegraphics[width=8.4cm, clip=True]{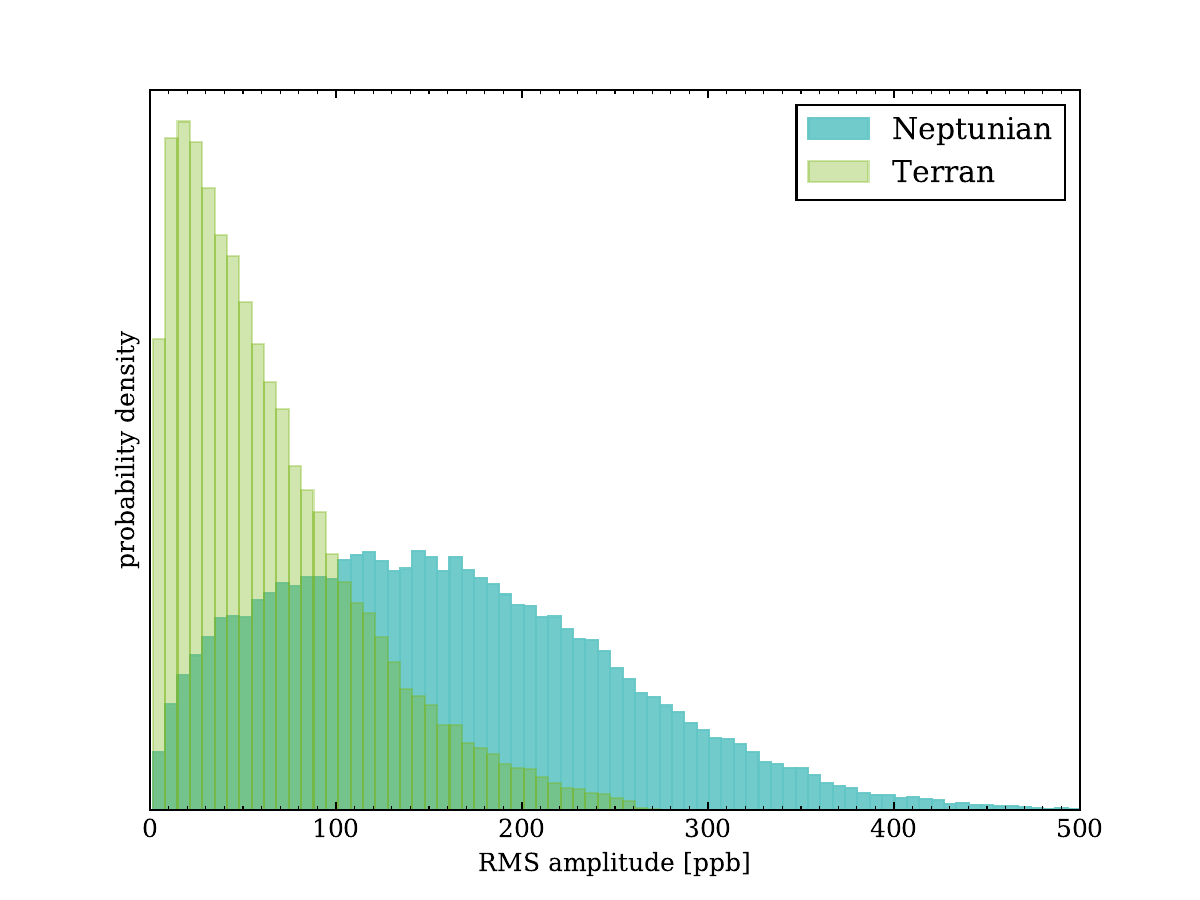}
\caption{
Posterior distributions of the root mean square amplitudes for the Neptunian
and Terran ensemble data. The fit to the Terran ensemble returns an upper limit
on the RMS amplitude of 169 parts per billion to the 95\% confidence level,
while the fit to the Neptunian ensemble returns an RMS amplitude of
$150^{+100}_{-90}$ parts per billion.
}
\label{fig:rms_posteriors}
\end{figure}

The RMS amplitude posterior for the Terran worlds, shown in
Figure~\ref{fig:rms_posteriors}, shows no strong offset from zero and thus
appears compatible with a null detection. Taking the median and surrounding
68.3\% quantile, the amplitude is measured to be $53_{-40}^{+60}$\,parts per
\textit{billion} - a remarkably precise photometric precision even by
\kepler's standards. As a result, we treat the albedo and greenhouse factor,
model parameters which are degenerate yet control amplitude, as also being null
detections. These null detections are corroborated by peaks in the posterior distributions at $A_{B} = 0$ and $f=1$. 

From our posteriors we instead measure an upper limit on the Bond
albedo of $A_{B} < 0.63$ to 95\% confidence, which under the Lambertian
assumption of our model, corresponds to an upper limit on the geometric albedo
of $A_{g} < 0.42$. Similarly, we measure an upper limit on the greenhouse
factor of $f < 1.6$ to 95\% confidence, and we find that the thermal
redistribution factor remains unconstrained. As expected, the vertical offset
$\gamma$ and error correction $\sigma_{\mathrm{jitter}}$ terms are marginal.
The posterior distributions for the fit to the model can be seen in 
Figure~\ref{fig:full-terran-corner}.

The posterior distribution of $\epsilon$ displays a non-uniform shape despite
using a uniform prior. The complex interplay of this parameter with the other
model parameters meant we did not have full conviction that this posterior was
not simply a general artifact of null detections. To test this, we scrambled
the Terran phase curve data randomly in phase, and then re-fitted using the
same algorithm. These posteriors, shown as black dashed lines in
Figure~\ref{fig:full-terran-corner}, reveal that indeed the $\epsilon$
posterior is an expected product of null detections and we give it little
weight as a significant effect in what follows.

We briefly highlight that the posteriors from the real data display a more
defined peak at $A_{B} =0$ than the scrambled data. Since the real data must
contain a genuine signal at some level (even if formally undetectable), it
is more likely to be coherent than scrambled data, potentially explaining
this observation. However, it is not exactly clear why the scrambled data and the real data do not produce identical posteriors, but it is also not obvious what level of difference should be interpreted as meaningful either. Nonetheless, it does not affect the final interpretation that there is no compelling case for a detection.
 
As described in Section~\ref{sub:SDDR}, we computed the Bayes factor between
the full model and a simpler black-planet model where we fix $A_B=0$.
This reveals a slight preference for the simpler model than the more general
case ($B_{\text{black:full}} = 3.35$), which is consistent with the result
expected from a null result.

All of the measured and derived values for the fit to the Terran ensemble data
can be seen in Table~\ref{table:terr-values}.

\def\arraystretch{1.5}
\begin{table}
\centering
\caption{
\textbf{Terran results:}
68.3\% credible intervals on the posterior distributions for our model
parameters. Upper panel lists the actual model parameters used, whereas
lower panel lists three other terms of interest.
}
\label{table:terr-values}
\begin{tabular}{lrr}
\toprule
\multicolumn{1}{c}{parameter} & \multicolumn{1}{c}{real data} & \multicolumn{1}{c}{scrambled data} \\ \midrule
$A_{B}$ & $0.18^{+0.24}_{-0.13}$ & $0.3^{+0.3}_{-0.2}$ \\
$f$ & $1.3^{+0.2}_{-0.2}$ & $1.3^{+0.3}_{-0.2}$ \\
$\epsilon$ & $-1^{+40}_{-30}$ & $-5^{+40}_{-30}$ \\
$\gamma$ [ppm] & $0.01^{+0.05}_{-0.01}$ & $0.01^{+0.04}_{-0.01}$ \\
$\sigma_{\text{jitter}}$ [ppm] & $0.47^{+0.08}_{-0.11}$ & $0.5^{+0.1}_{-0.1}$\\
\midrule
$A_{g}$ & $0.12^{+0.16}_{-0.09}$ & $0.20^{+0.22}_{-0.14}$\\
RMS [ppb] & $53^{+60}_{-40}$ & $86^{+90}_{-60}$\\
$B(A_B=0:A_B>0)$ & $3.35$ & $1.96$ \\
\bottomrule
\end{tabular}
\end{table}

\def\arraystretch{1.5}
\begin{table}
\centering
\caption{
\textbf{Neptunian results:}
68.3\% credible intervals on the posterior distributions for our model
parameters. Upper panel lists the actual model parameters used, whereas
lower panel lists three other terms of interest.
}
\label{table:nep-values}
\begin{tabular}{lrr}
\toprule
\multicolumn{1}{c}{parameter} & \multicolumn{1}{c}{real data} & \multicolumn{1}{c}{scrambled data} \\ \midrule
$A_{B}$ & $0.15^{+0.12}_{-0.10}$ & $0.08^{+0.1}_{-0.06}$ \\
$f$ & $1.18^{+0.14}_{-0.12}$ & $1.18^{+0.14}_{-0.12}$ \\
$\epsilon$ & $-6^{+40}_{-30}$ & $-7^{+40}_{-30}$ \\
$\gamma$ [ppm] & $0.01^{+0.05}_{-0.01}$ & $0.01^{+0.06}_{-0.01}$ \\
$\sigma_{\text{jitter}}$ [ppm] & $0.14^{+0.23}_{-0.14}$ & $0.17^{+0.21}_{-0.17}$\\
\midrule
$A_{g}$ & $0.10^{+0.08}_{-0.07}$ & $0.06^{+0.07}_{-0.04}$\\
RMS [ppb] & $150^{+100}_{-90}$ & $87^{+90}_{-60}$\\
$B(A_B=0:A_B>0)$ & $2.77$ & $6.46$ \\
\bottomrule
\end{tabular}
\end{table}

\subsection{Neptunian ensemble}

For the Neptunian ensemble, we again computed a posterior distribution for
the RMS phase curve amplitude and find it peaks somewhat away from zero, as
can be seen in Figure~\ref{fig:rms_posteriors}, giving $150^{+100}_{-90}$\,ppb.
To estimate the significance, we use the \citet{lucy:1971} test to derive
a false alarm probability of 5.8\% (1.9\,$\sigma$). Accordingly, whilst
certainly intriguing, we do not consider this to be a significant
``detection'' and thus treat the inferred albedo and greenhouse factor
parameters as upper limits, as with the Terran sample.

From the marginalized posteriors, we derived an upper limit on the ensemble
Neptunian Bond albedo of $A_{B} < 0.35$ to 95\% confidence, or $A_g < 0.23$
when converted to a geometric albedo. As before, the thermal redistribution
efficiency remains unconstrained but we can constrain the greenhouse factor
to be $f < 1.40$ to 95\% confidence. The posterior distributions of the fit to
the full thermal-reflection model can be seen in 
Figure~\ref{fig:full-neptunian-corner}. The vertical offset $\gamma$ and error
correction $\sigma_{jitter}$ terms are again marginal.

Comparing the posterior distributions of the full fit to the data to that of
the fit to the scrambled in Figure~\ref{fig:full-neptunian-corner} shows that
they are nearly identical for all parameters except for the Bond albedo.
Here, the scrambled data produces a posterior peaked at zero whereas the
real data prefer a slightly positive value, consistent with tentative evidence
for a detection found from the RMS amplitude posterior. Despite the posterior
distribution of the thermal redistribution efficiency showing a preference for
negative values (i.e. subrotating winds), this behavior is also displayed by
the fit to the scrambled data, which again suggests that this is not significant.

The Bayes factor as estimated by the Savage-Dickey ratio for the true Neptunian
ensemble data is determined to be $B_{\text{black:full}} = 2.77$, which
suggests that there is no substantial evidence for a preference for the
black-planet model over the full model. The weak preference of the black-planet model over the full model can be
corresponded to a 1.1\,$\sigma$ significance
($\sqrt{2}\mathrm{erfc}^{-1}[1/(B+1)]$; see \citealt{hatp24}). We note however that the degree to
which the black planet model is favored is less than that of the scrambled
data, the opposite to what happened with the definitively null detection of the
Terran ensemble. 

All of the measured and derived values for the fit to the Neptunian ensemble
data can be seen in Table \ref{table:nep-values}.

We briefly comment that inspection of Figure~\ref{fig:full-neptunian-corner}
reveals a suggestive transit-like feature at around $-0.25$ phase. We conducted
a nonlinear least squares regression of a box-like dip seeding from around
this phase and achieved a $\chi^2$ improvement of 14.1 versus a flat-line.
Given that three extra free parameters are necessary (a mid-time, a depth
and a duration), this improvement does not outweigh the likelihood penalty
from the \citet{schwarz:1978} BIC criterion ($3 \log 450 = 18.3$) and thus
we do not consider it to be significant.

To ensure this feature does not significantly skew our results, we removed
the points between phases $-0.26$ and $-0.18$ and repeated our fits from
earlier. Referencing the results in Table \ref{table:nep-minus-phase-values}, we find that the maximum a-posteriori values change by less than
one-sigma and thus this feature does not appear to significantly affect our
inferences. Naturally, when conditioning our inferences upon this truncated data set the upper limits described earlier tend to be slightly modified, with a 95\% confidence upper limit on the Bond albedo of $A_{B}<0.26$, or $A_{g}<0.17$ when converted to geometric albedo. As with the full set of data, the thermal redistribution efficiency remains unconstrained, while an upper limit on the greenhouse factor is constrained to $f<1.47$ to 95\% confidence.

\def\arraystretch{1.5}
\begin{table}
\centering
\caption{
\textbf{Results of the Neptunian phase curve minus points of interest:}
68.3\% credible intervals on the posterior distributions for our model
parameters on the Neptunian ensemble with points between phases -0.26 and -0.18 removed. Upper panel lists the actual model parameters used, whereas
lower panel lists three other terms of interest. 
}
\label{table:nep-minus-phase-values}
\begin{tabular}{lrr}
\toprule
\multicolumn{1}{c}{parameter} & \multicolumn{1}{c}{real data} & \multicolumn{1}{c}{scrambled data} \\ \midrule
$A_{B}$ & $0.07^{+0.1}_{-0.06}$ & $0.11^{+0.07}_{-0.09}$ \\
$f$ & $1.26^{+0.15}_{-0.18}$ & $1.25^{+0.15}_{-0.16}$\\
$\epsilon$ & $8^{+30}_{-40}$ &$-5^{+40}_{-30}$ \\
$\gamma$ [ppm] & $0.02^{+0.14}_{-0.02}$ & $0.02^{+0.13}_{-0.02}$\\
$\sigma_{\text{jitter}}$ [ppm] & $0.09^{+0.3}_{-0.09}$ & $0.09^{+0.3}_{-0.04}$\\
\midrule
$A_{g}$ & $0.05^{+0.07}_{-0.04}$ & $0.07^{+0.05}_{-0.06}$ \\
RMS [ppb] & $78^{+80}_{-50}$ & $114^{+100}_{-70}$ \\
\bottomrule
\end{tabular}
\end{table}

%% file: discussion.tex
In this work, we have produced the first demonstration of a population-stack
of exoplanet phase curves for 50 Terran and 115 Neptunian confirmed \kepler\
planets. Enabling this analysis, we have devised a simple but powerful
non-parametric detrending algorithm optimized for reconstructing the phase
curves of transiting planets, dubbed \phasma. A mathematical motivation,
example test, and a battery of control cases are presented to justify our
use of this new code.

After detrending, stacking and regressing ensemble phase curve models to our
data product, we find modest evidence (2.4\,$\sigma$) for a coherent signal in
the 115-planet Neptunian sample, with an R.M.S. amplitude of $150_{-90}^{+100}$
parts per \textit{billion}. Given the weak significance of the signal and the
covariances between atmospheric model parameters, the model parameters have
greater fractional uncertainty, with the 68.3\% credible interval on geometric
albedo being $A_g = 0.10_{-0.07}^{+0.08}$ for the Neptunian ensemble. This measurement peaks slightly
further out than our fits on control-data without any signal, but we do not
consider it significant enough to claim a detection. Accordingly, we use the
posterior to derive an upper limit of $A_g<0.23$ to 95\% confidence.

This result indicates that most of the Neptunians in our ensemble are
significantly darker than Neptune, which has a full-disk albedo of 0.30
when integrated across the \kepler\ response function (using the albedo
spectrum from \citealt{karkoschka:1994}). Lower albedos for warm Neptunians
has been predicted theoretically, as a consequence of temperatures becoming
too warm for bright clouds to form \citep{cahoy:2010}. The increased insolation
on these short-period planets can also lead to Doppler broadened spectral
lines, for example of sodium and potassium, which can also contribute to a lower
albedo \citep{sudarsky:2000, seager:2000b, spiegel:2010, heng:2016}.

Our result may be compared to the stacked occultation measurement of
\citet{sheets:2017}, who find a typical geometric albedo of
$A_g = (0.05 \pm 0.04)$ and $(0.11 \pm 0.06)$ for $2-4$\,$R_{\oplus}$ and
$4-6$\,$R_{\oplus}$ \kepler\ planets, respectively. However,
there are several important differences in the samples (some of these can be
seen illustratively in Figure~\ref{fig:targets}). First, our definition
of a Neptunian planet, deduced by the classification algorithm \forecaster,
does not directly correspond to the radius cut of $2-6$\,$R_{\oplus}$, rather
we extend down further to $\sim 1.25$\,$R_{\oplus}$.
Second, the Neptunian planets used in this work are all dispositioned as
``CONFIRMED'' rather than ``CANDIDATE'' on the NASA Exoplanet Archive. Finally,
our sample is considerably cooler than the planets used by \citet{sheets:2017},
whose sample has a mean equilibrium temperature of $\sim 2000$\,K versus
$\sim 1200$\,K for this work. Nevertheless, the results are broadly consistent
and support a warm Neptunian albedo of $A_g \lesssim 0.2$.

For the 50-planet Terran sample, we have derived what appears to be the first
measurement of this type of planet's average albedo. Whilst \citet{sheets:2017}
do have a category dubbed ``Super-Earths'', spanning $1-2$\,$R_{\oplus}$,
probabilistic classifications from \forecaster\ \citep{chen:2017a} indicate
that the majority of such planets are more likely to belong to a population with a
mass-radius relation describing gaseous bodies, rather than solid ones. Aside
from the predictions of \forecaster, we highlight that other independent
studies support the argument that a $1-2$\,$R_{\oplus}$ category would, at the
very least, have significant contamination of mini-Neptunes within its sample
(e.g. see \citealt{lopez:2014,rogers:2015,lehmer:2017,fulton:2017}).

Although our work derives the first Terran-ensemble albedo, the measurement
is an upper limit rather than a detection, as with the Neptunian set.
Specifically, we measure an upper limit on the representative geometric albedo
of our 50-planet Terran ensemble to be $A_g<0.42$ to 95\% confidence. This
excludes a Venusian geometric albedo of 0.67 (i.e. planets covered in
reflective clouds or hazes), as well as the less likely case of icy-covered
surface such as Europa or Enceladus. However our measurement remains compatible
with a Mercurian or Martian value (0.14 and 0.17 respectively; albedos taken
from \citealt{mallama:2009}).

With a mean temperature of $\sim 1000$\,K, the Terrans considered here are
unlikely to be covered in lava-oceans, as has been hypothesized for ultra-short
period \kepler\ planets \citep{rouan:2011} and so these worlds are more likely
to resemble Mercury. Aside from an albedo constraint, we also find that the
null detection constrains the thermal component such that the marginalized
greenhouse factor must be $f<1.60$ to the 95\% confidence level. This result
is inconsistent with a Venusian strong greenhouse, which together with the
lower albedo adds weight to contention that our sample of Terran planets are
likely non-Venusian in nature. This would be compatible with a lack of
thick atmosphere as a product of photo-evaporative sculpting
\citep{lehmer:2017}, leaving behind a dark, basaltic surface \citep{hu:2012}.

Continued photometric surveys for transiting planets promises to greatly
increase our sample of small planets suitable for analysis. The techniques
described in this work show great promise to measure the albedo of larger
samples of planets, even at longer orbital periods in the near future.

\begin{figure*}
\includegraphics[width=15cm]{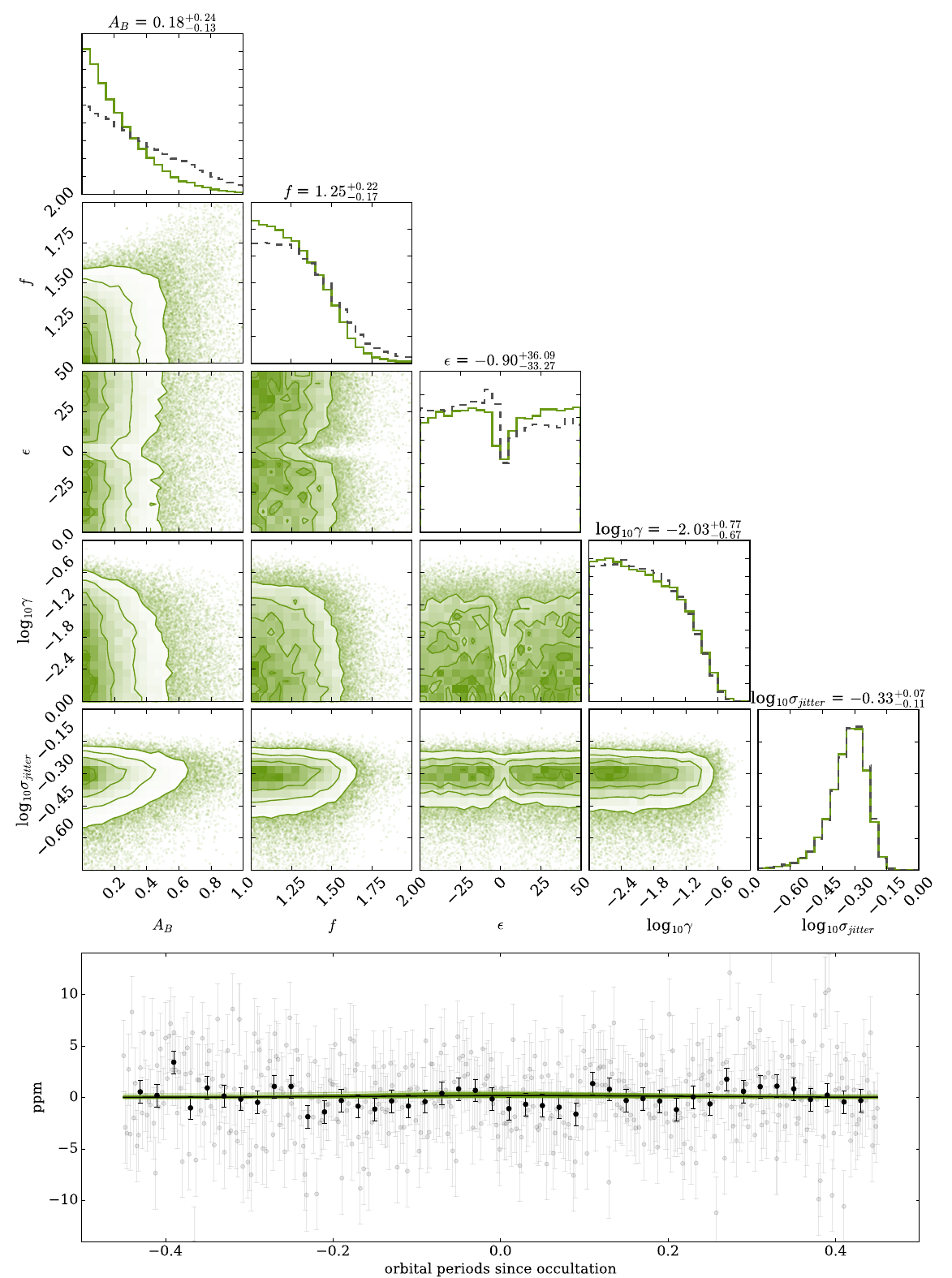}
\caption{Posterior distributions of the full thermal-reflection model fit for
the Terran ensemble, where the 1-D histograms show the probability densities
of the fit to the ensemble data in solid-green and to the flat-line data in dashed-grey
[top]. The bottom panel shows the binned ensemble data (500 bins in grey, 50
bins in black) plotted with model phase curves constructed from parameters
sampled from the posterior distributions (green), where the median phase curve
is overplotted in black.
\label{fig:full-terran-corner}}
\end{figure*}

\begin{figure*}
\includegraphics[width=15cm]{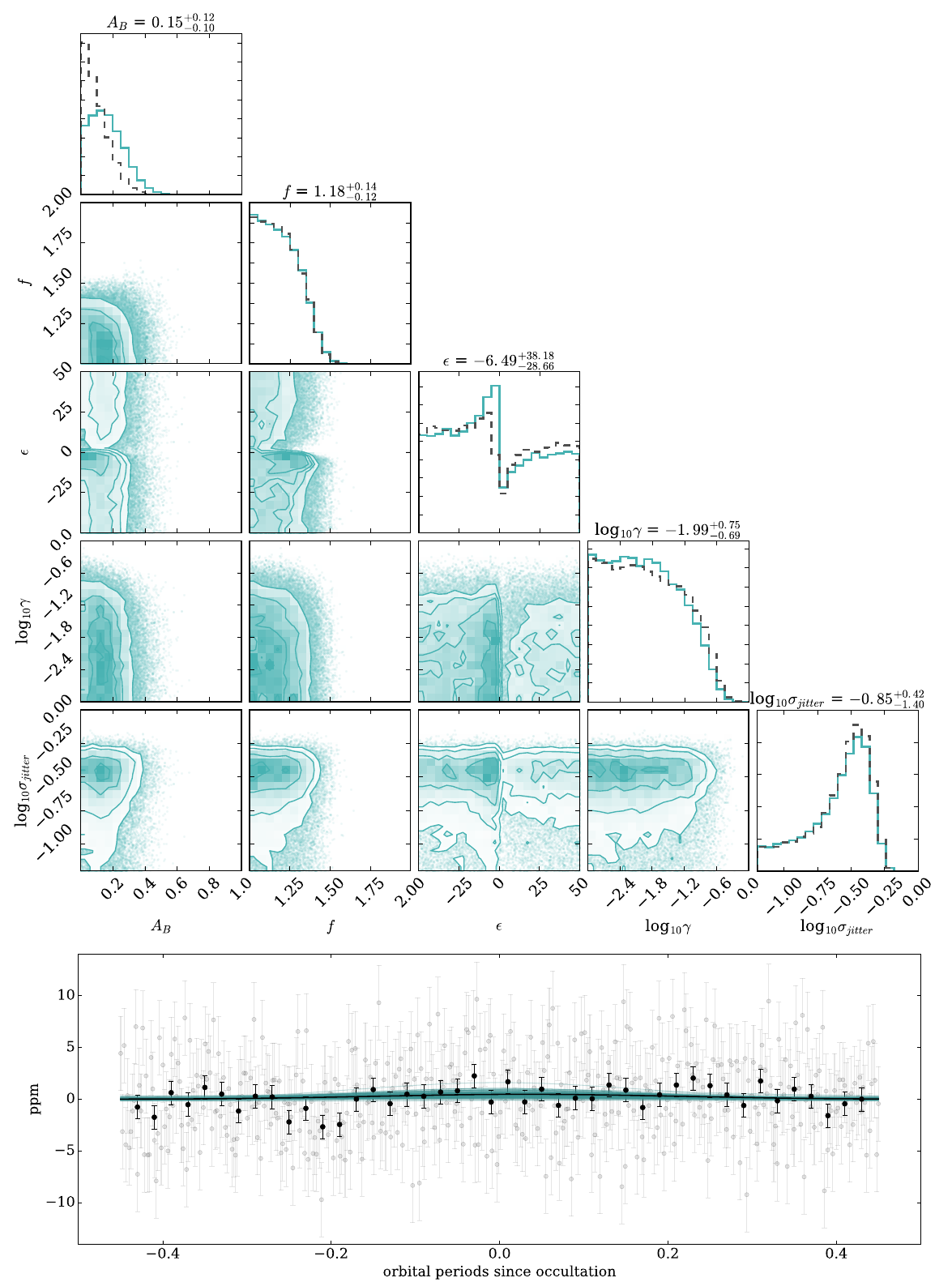}
\caption{Posterior distributions of the full thermal-reflection model fit for
the Neptunian ensemble, where the 1-D histograms show the probability densities
of the fit to the ensemble data in solid-blue and to the flat-line data in dashed-grey [top].
The bottom panel shows the binned ensemble data (500 bins in grey, 50 bins in
black) plotted with model phase curves constructed from parameters sampled from
the posterior distributions (blue), where the median phase curve is overplotted
in black.
\label{fig:full-neptunian-corner}}
\end{figure*}

%% file: longtables.tex
\begin{table}
\centering
\caption{KOIs used in the Terran ensemble.}
\label{table:terr_KOIs}
\renewcommand{\arraystretch}{0.95}
\begin{tabular}{lrr}
\toprule
KOI & \multicolumn{1}{c}{period {[}days{]}} & \multicolumn{1}{c}{planetary radius {[}$R_{\oplus}${]}} \\
\hline
K02071.01 & $3.86$ & $1.03$ \\
K04230.01 & $5.50$ & $1.13$ \\
K01612.01 & $2.47$ & $0.67$ \\
K01536.01 & $3.74$ & $1.07$ \\
K05827.01 & $6.11$ & $0.70$ \\
K02208.01 & $2.34$ & $0.86$ \\
K02317.01 & $3.79$ & $1.19$ \\
K01002.01 & $3.48$ & $1.16$ \\
K04312.01 & $7.85$ & $0.75$ \\
K04585.01 & $6.76$ & $1.10$ \\
K02260.01 & $6.12$ & $1.01$ \\
K02832.01 & $3.58$ & $0.84$ \\
K04232.01 & $2.62$ & $1.20$ \\
K02372.01 & $5.35$ & $1.18$ \\
K04360.01 & $2.72$ & $1.12$ \\
K03015.01 & $3.61$ & $0.85$ \\
K02877.01 & $5.31$ & $1.21$ \\
K01434.01 & $2.34$ & $1.22$ \\
K03892.01 & $2.42$ & $0.99$ \\
K02730.01 & $4.52$ & $1.22$ \\
K04292.01 & $9.33$ & $0.61$ \\
K03141.01 & $2.32$ & $1.20$ \\
K04190.01 & $3.43$ & $0.80$ \\
K02561.01 & $3.24$ & $0.86$ \\
K02913.01 & $2.89$ & $0.82$ \\
K02143.01 & $4.79$ & $1.10$ \\
K02951.01 & $2.44$ & $0.72$ \\
K04347.01 & $3.06$ & $1.14$ \\
K01967.01 & $4.42$ & $1.15$ \\
K04513.01 & $3.92$ & $0.99$ \\
K02426.01 & $4.16$ & $0.74$ \\
K02509.01 & $4.55$ & $1.09$ \\
K04117.01 & $4.24$ & $1.17$ \\
K01202.01 & $0.93$ & $1.23$ \\
K04510.01 & $5.18$ & $1.25$ \\
K03113.01 & $2.46$ & $1.04$ \\
K02742.01 & $0.79$ & $0.95$ \\
K02798.01 & $0.92$ & $0.61$ \\
K01528.01 & $3.99$ & $1.02$ \\
K02344.01 & $1.12$ & $1.08$ \\
K02247.01 & $4.46$ & $0.82$ \\
K02399.01 & $1.92$ & $0.86$ \\
K02101.01 & $2.89$ & $1.17$ \\
K02347.01 & $0.59$ & $0.98$ \\
K01300.01 & $0.63$ & $1.11$ \\
K02845.01 & $1.57$ & $1.08$ \\
K02058.01 & $1.52$ & $0.98$ \\
K02238.01 & $1.65$ & $0.88$ \\
K04268.01 & $0.85$ & $0.58$ \\
K02662.01 & $2.10$ & $0.73$ \\
\hline
\end{tabular}
\end{table}

\begin{table*}
\centering
\caption{KOIs used in the Neptunian ensemble.}
\label{table:nep_KOIs}
\renewcommand{\arraystretch}{0.95}
\begin{tabular}{lrr}
\toprule
\multicolumn{1}{c}{KOI} & \multicolumn{1}{c}{period {[}days{]}} & \multicolumn{1}{c}{planetary radius {[}R$_{\oplus}${]}} \\
\hline
K00697.01 & $3.03$ & $3.01$ \\
K00007.01 & $3.21$ & $4.21$ \\
K01890.01 & $4.34$ & $1.69$ \\
K00155.01 & $5.66$ & $2.99$ \\
K01632.01 & $4.59$ & $1.38$ \\
K00265.01 & $3.57$ & $1.28$ \\
K01116.01 & $3.75$ & $1.40$ \\
K01339.01 & $4.17$ & $2.02$ \\
K01214.01 & $4.24$ & $1.40$ \\
K01341.01 & $4.51$ & $1.85$ \\
K01629.01 & $4.41$ & $1.38$ \\
K00578.01 & $6.41$ & $4.00$ \\
K02995.01 & $5.19$ & $2.51$ \\
K02785.01 & $4.77$ & $1.97$ \\
K00167.01 & $4.92$ & $2.80$ \\
K02437.01 & $4.15$ & $1.79$ \\
K02552.01 & $3.66$ & $1.36$ \\
K00452.01 & $3.71$ & $2.55$ \\
K02331.01 & $2.83$ & $1.30$ \\
K02440.01 & $4.87$ & $1.31$ \\
K00292.01 & $2.59$ & $1.43$ \\
K03120.01 & $4.14$ & $1.49$ \\
K00955.01 & $7.04$ & $2.87$ \\
K02778.01 & $2.22$ & $1.64$ \\
K02200.01 & $3.17$ & $1.68$ \\
K01526.01 & $4.44$ & $1.64$ \\
K00240.01 & $4.29$ & $4.47$ \\
K03110.01 & $4.48$ & $1.37$ \\
K00535.01 & $5.85$ & $3.87$ \\
K04139.01 & $3.74$ & $1.56$ \\
K00893.01 & $4.41$ & $2.59$ \\
K01491.01 & $3.15$ & $2.10$ \\
K02780.01 & $3.35$ & $2.37$ \\
K01570.01 & $6.34$ & $3.31$ \\
K02016.01 & $4.31$ & $1.99$ \\
K00069.01 & $4.73$ & $1.48$ \\
K00926.01 & $3.17$ & $4.43$ \\
K00852.01 & $3.76$ & $2.40$ \\
K01603.01 & $3.02$ & $1.49$ \\
K01706.01 & $1.92$ & $1.40$ \\
K00532.01 & $4.22$ & $2.93$ \\
K00714.01 & $4.18$ & $2.63$ \\
K02383.01 & $4.37$ & $1.62$ \\
K01979.01 & $2.71$ & $1.30$ \\
K01481.01 & $5.10$ & $3.22$ \\
K00916.01 & $3.31$ & $3.94$ \\
K02300.01 & $2.69$ & $1.31$ \\
K00762.01 & $4.50$ & $2.49$ \\
K01641.01 & $4.85$ & $1.52$ \\
K00537.01 & $2.82$ & $2.48$ \\
K00794.01 & $2.54$ & $2.10$ \\
K00583.01 & $2.44$ & $1.91$ \\
K00161.01 & $3.11$ & $2.48$ \\
K01304.01 & $4.60$ & $1.98$ \\
K02615.01 & $4.70$ & $1.70$ \\
K05466.01 & $4.10$ & $1.55$ \\
K00863.01 & $3.17$ & $3.15$ \\
K00361.01 & $3.25$ & $1.51$ \\
\multicolumn{1}{c}{continued to the right...}\\
\end{tabular}
\begin{tabular}{lrr}
\multicolumn{1}{c}{KOI} & \multicolumn{1}{c}{period {[}days{]}} & \multicolumn{1}{c}{planetary radius {[}R$_{\oplus}${]}} \\
\hline
K02619.01 & $3.28$ & $2.17$ \\
K01094.01 & $6.10$ & $3.16$ \\
K01885.01 & $5.65$ & $2.13$ \\
K02146.01 & $2.95$ & $1.49$ \\
K02789.01 & $3.40$ & $1.36$ \\
K00786.01 & $3.69$ & $2.21$ \\
K01505.01 & $5.03$ & $2.27$ \\
K02213.01 & $3.97$ & $1.42$ \\
K02100.01 & $4.27$ & $1.42$ \\
K04241.01 & $2.55$ & $1.47$ \\
K01488.01 & $3.95$ & $2.42$ \\
K02155.01 & $4.34$ & $1.80$ \\
K02491.01 & $2.67$ & $1.72$ \\
K01428.01 & $0.93$ & $1.70$ \\
K00923.01 & $5.74$ & $3.65$ \\
K01626.01 & $2.53$ & $1.87$ \\
K02223.01 & $1.10$ & $1.64$ \\
K02355.01 & $1.22$ & $1.62$ \\
K01511.01 & $2.58$ & $1.81$ \\
K02266.01 & $1.00$ & $1.69$ \\
K03848.01 & $1.85$ & $1.69$ \\
K00826.01 & $6.37$ & $2.74$ \\
K00483.01 & $4.80$ & $2.75$ \\
K01965.01 & $2.51$ & $1.44$ \\
K00769.01 & $4.28$ & $2.58$ \\
K00844.01 & $3.71$ & $4.79$ \\
K00472.01 & $4.24$ & $3.64$ \\
K02580.01 & $3.12$ & $1.87$ \\
K01337.01 & $1.92$ & $1.37$ \\
K01882.01 & $3.77$ & $2.08$ \\
K00585.01 & $3.72$ & $2.50$ \\
K01501.01 & $2.62$ & $1.64$ \\
K01893.01 & $3.55$ & $1.48$ \\
K00144.01 & $4.18$ & $3.08$ \\
K04566.01 & $3.95$ & $1.65$ \\
K02820.01 & $3.01$ & $1.37$ \\
K01344.01 & $4.49$ & $1.30$ \\
K00943.01 & $3.60$ & $2.38$ \\
K02063.01 & $3.01$ & $1.49$ \\
K00910.01 & $5.39$ & $2.90$ \\
K01142.01 & $3.76$ & $1.81$ \\
K00873.01 & $4.35$ & $2.34$ \\
K02214.01 & $2.35$ & $1.40$ \\
K01637.01 & $2.97$ & $1.92$ \\
K00861.01 & $2.24$ & $1.54$ \\
K02708.01 & $0.87$ & $1.71$ \\
K01424.01 & $1.22$ & $1.64$ \\
K01973.01 & $3.29$ & $1.55$ \\
K00104.01 & $2.51$ & $3.15$ \\
K02839.01 & $2.16$ & $1.34$ \\
K01880.01 & $1.15$ & $1.36$ \\
K01577.01 & $2.81$ & $1.49$ \\
K00739.01 & $1.29$ & $1.45$ \\
K02156.01 & $2.85$ & $1.75$ \\
K00778.01 & $2.24$ & $1.69$ \\
K02705.01 & $2.89$ & $1.34$ \\
K04928.01 & $3.29$ & $5.47$ \\
\hline
\end{tabular}
\end{table*}

%% file: manuscript.bbl
\begin{thebibliography}{99}
\bibitem[\protect\citeauthoryear{Akeson et al.}{2013}]{akeson:2013} 
Akeson, R,~L., Chen, X., Ciardi, D., et al. 2013, PASP, 125, 989
\bibitem[\protect\citeauthoryear{Angerhausen et al.}{2015}]{angerhausen:2015}
Angerhausen, D., DeLarme, E., Morse, J.~A., 2015, PASP, 127, 1113
\bibitem[\protect\citeauthoryear{Armstrong et al.}{2016}]{armstrong:2016} 
Armstrong, D.~J., de Mooij, E., Barstow, J., Osborn, H.~P., Blake, J. \& Fereshteh
Saniee, N. 2016, Nature Astronomy, 1, 4
\bibitem[\protect\citeauthoryear{Barclay et al.}{2012}]{barclay:2012} 
Barclay, T., Huber, D., Rowe, J.~F., Fortney, J.~J., Morley, C.~V.,
Quintana, E.~V., Fabrycky, D.~C., Barentsen, G., Bloemen, S.,
Christiansen, J.~L., Demory, B.-O., Fulton, B.~J., Jenkins, J.~M.,
Mullally, F., Ragozzine, D., Seader, S.~E., Shporer, A., Tenenbaum, P.
\& Thompson, S.~E., 2012, ApJ, 761, 53
\bibitem[\protect\citeauthoryear{Batalha et al.}{2011}]{batalha:2011} 
Batalha, N.~M., Borucki, W.~J., Bryson, S.~T., Buchhave, L.~A.,
Caldwell, D.~A., Christensen-Dalsgaard, J., Ciardi, D., Dunham, E.~W.,
Fressin, F., Gautier, T.~N., Gilliland, R.~L., Haas, M.~R., Howell, S.~B.,
Jenkins, J.~M., Kjeldsen, H., Koch, D.~G., Latham, D.~W., Lissauer, J.~J.,
Marcy, G.~W., Rowe, J.~F., Sasselov, D.~D., Seager, S., Steffen, J.~H.,
Torres, G., Basri, G.~S., Brown, T.~M., Charbonneau, D., Christiansen, J.,
Clarke, B., Cochran, W.~D., Dupree, A., Fabrycky, D.~C., Fischer, D.,
Ford, E.~B., Fortney, J., Girouard, F.~R., Holman, M.~J., Johnson, J.~A.,
Isaacson, H., Klaus, T.~C., Machalek, P., Moorehead, A.~V., Morehead, R.~C.,
Ragozzine, D., Tenenbaum, P., Twicken, J., Quinn, S., VanCleve, J.,
Walkowicz, L.~M., Welsh, W.~F., Devore, E. \& Gould, A., 2011, ApJ, 729, 27
\bibitem[\protect\citeauthoryear{Birkby et al.}{2013}]{birkby:2013} 
Birkby, J.~L., de Kok, R.~J., Brogi, M., de Mooij, E.~J.~W., Schwarz, H.,
Albrecht, S. \& Snellen, I.~A.~G., 2013, MNRAS, 436, 35
\bibitem[\protect\citeauthoryear{Borucki et al.}{2009}]{borucki:2009} 
Borucki, W.~J., Koch, D., Jenkins, J., et al.\ 2009, Science, 325, 709
\bibitem[\protect\citeauthoryear{Burrows}{2014}]{burrows:2014} 
Burrows, A.~S., 2014, Nature, 513, 345
\bibitem[\protect\citeauthoryear{Cahoy et al.}{2010}]{cahoy:2010}
Cahoy, K.~L., Marley, M.~S. \& Fortney, J.~J. 2010, ApJ, 724, 189
\bibitem[\protect\citeauthoryear{Charbonneau et al.}{2002}]{charbonneau:2002}
Charbonneau, D., Brown, T.~M., Noyes, R.~W. \& Gilliland, R.~L.  2002,
ApJ, 568, 377
\bibitem[\protect\citeauthoryear{Charbonneau et al.}{2008}]{charbonneau:2008}
Charbonneau, D., Knutson, H.~A., Barman, T., Allen, L.~E., Mayor, M.,
Megeath, S.~T., Queloz, D. \& Udry, S., 2008, ApJ, 686, 1341
\bibitem[\protect\citeauthoryear{Chen et al.}{2017}]{chen:2017a} 
Chen, J. \& Kipping, D.~M. 2017, ApJ, 834, 17
\bibitem[\protect\citeauthoryear{Chen et al.}{2018}]{chen:2017b} 
Chen, J. \& Kipping, D.~M. 2018, MNRAS, 473, 2753  
\bibitem[\protect\citeauthoryear{Christiansen et al.}{2013}]{christiansen:2013} 
Christiansen, J.~L., Jenkins, J.~M., Caldwell, D.~A., Burke, C.~J.,
Tenenbaum, P., Seader, S., Thompson, S.~E., Barclay, T.~S., Clarke, B.~D.,
Li, J., Smith, J.~C., Stumpe, M.~C., Twicken, J.~D., Van Cleve, J., 2013,
PASP, 124, 1279
\bibitem[\protect\citeauthoryear{Claret \& Bloemen}{2011}]{claret:2011} 
Claret, A. \& Bloemen, S. 2011, A\&A, 529, 75
\bibitem[\protect\citeauthoryear{Cowan \& Agol}{2011}]{cowan:2011a}
Cowan, N.~B., \& Agol, E.\ 2011, ApJ, 726, 82 
\bibitem[\protect\citeauthoryear{Cowan \& Agol}{2011b}]{cowan:2011b} 
Cowan, N.~B., \& Agol, E. 2011b, ApJ, 729, 54 
\bibitem[\protect\citeauthoryear{Deleuil et al.}{2014}]{deleuil:2014} 
Deleuil, M., Almenara, J.-M., Santerne, A., Barros, S.~C.~C., Havel, M.,
H\'ebrard, G., Bonomo, A.~S., Bouchy, F., Bruno, G., Damiani, C.,
D\'iaz, R.~F., Montagnier, G. \& Moutou, C. 2014, A\&A, 564, 56
\bibitem[\protect\citeauthoryear{Deming et al.}{2005}]{deming:2005}
Deming, D., Seager, S., Richardson, L.~J., \& Harrington, J. 2005,
Nature, 434, 740
\bibitem[\protect\citeauthoryear{Demory et al.}{2011}]{demory:2011}
Demory, B.-O., Seager, S., Madhusudhan, N., Kjeldsen, H.,
Christensen-Dalsgaard, J., Gillon, M., Rowe, J.~F., Welsh, W.~F., Adams, E.~R.,
Dupree, A., McCarthy, D., Kulesa, C., Borucki, W.~J. \& Koch, D.~G. 2011,
ApJ, 735, 12
\bibitem[\protect\citeauthoryear{Demory et al.}{2013}]{demory:2013}
Demory, B.-O., de Wit, J., Lewis, N., Fortney, J., Zsom, A., Seager, S.,
Knutson, H., Heng, K., Madhusudhan, N., Gillon, M., Barclay, T.,
Desert, J.-M., Parmentier, V. \& Cowan, N.~B., 2013, ApJ, 776, 25
\bibitem[\protect\citeauthoryear{Demory}{2014}]{demory:2014} 
Demory, B.-O. 2014, ApJ, 789, 20
\bibitem[\protect\citeauthoryear{D\'esert et al.}{2011}]{desert:2011}
D\'esert, J.-M., Charbonneau, D., Fortney, J.~J., Madhusudhan, N.,
Knutson, H.~A., Fressin, F., Deming, D., Borucki, W.~J., Brown, T.~M.,
Caldwell, D., Ford, E.~B., Gilliland, R.~L., Latham, D.~W., Marcy, G.~W.,
Seager, S. 2011, ApJS, 197, 11
\bibitem[\protect\citeauthoryear{Dickey \& Lientz}{1970}]{dickey:1970}
Dickey, J. M., \& Lientz, B.P. 1970, The Annals of Mathematical Statistics, 41, 214-226
\bibitem[\protect\citeauthoryear{Esteves et al.}{2013}]{esteves:2013}
Esteves, L.~J., De Mooij, E.~J.~W. \& Jayawardhana, R. 2013, ApJ, 772, 51
\bibitem[\protect\citeauthoryear{Foreman-Mackey et al.}{2013}]{dfm:2013} 
Foreman-Mackey, D., Hogg, D.~W., Lang, D. \& Goodman, J. 2013, PASP, 125, 306
\bibitem[\protect\citeauthoryear{Fortney et al.}{2011}]{fortney:2011} 
Fortney, J.~J., Demory, B.-O., D\'esert, J.-M., Rowe, J., Marcy, G.~W.,
Isaacson, H., Buchhave, L.~A., Ciardi, D., Gautier, T.~N., Batalha, N.~M.,
Caldwell, D.~A., Bryson, S.~T., Nutzman, P., Jenkins, J.~M., Howard, A.,
Charbonneau, D., Knutson, H.~A., Howell, S.~B., Everett, M., Fressin, F.,
Deming, D., Borucki, W.~J., Brown, T.~M., Ford, E.~B., Gilliland, R.~L.,
Latham, D.~W., Miller, N., Seager, S., Fischer, D.~A., Koch, D.,
Lissauer, J.~J., Haas, M.~R., Still, M., Lucas, P., Gillon, M.,
Christiansen, J.~L. \& Geary, J.~C. 2011, ApJS, 197, 9
\bibitem[\protect\citeauthoryear{Fulton et al.}{2017}]{fulton:2017} 
Fulton, B.~J., Petigura, E.~A., Howard, A.~W., Isaacson, H., Marcy, G.~W.,
Cargile, P.~A., Hebb, L., Weiss, L.~M., Johnson, J.~A., Morton, T.~D.,
Sinukoff, E., Crossfield, I.~J.~M. \& Hirsch, L.~A. 2017, AJ, 154, 109
\bibitem[\protect\citeauthoryear{Gandolfi et al.}{2015}]{gandolfi:2015} 
Gandolfi, D., Parviainen, H., Deeg, H.~J., Lanza, A.~F., Fridlund, M.,
Prada Moroni, P.~G., Alonso, R., Augusteijn, T., Cabrera, J., Evans, T.,
Geier, S., Hatzes, A.~P., Holczer, T., Hoyer, S., Kangas, T., Mazeh, T.,
Pagano, I., Tal-Or, L. \& Tingley, B. 2015, A\&A, 576, 11
\bibitem[\protect\citeauthoryear{Hansen}{2008}]{hansen:2008} 
Hansen, B.~M.~S. 2008, ApJS, 179, 484
\bibitem[\protect\citeauthoryear{Heng}{2016}]{heng:2016}
Heng, K. 2016, ApJL, 826, L16
\bibitem[\protect\citeauthoryear{Hippke \& Angerhausen}{2015}]{hippke:2015} 
Hippke, M. \& Angerhausen, D., 2015, ApJ, 811, 1
\bibitem[\protect\citeauthoryear{Hu et al.}{2012}]{hu:2012}
Hu, R., Ehlmann, B.~L., Seager, S., 2012, ApJ, 752, 7
\bibitem[\protect\citeauthoryear{Hu et al.}{2015}]{hu:2015}
Hu, R., Demory, B.-O., Seager, S., Lewis, N., \& Showman, A.~P.\ 2015,
ApJ, 802, 51
\bibitem[\protect\citeauthoryear{Huber}{1981}]{huber:1981}
Huber, P. J. 1981, \textit{Robust Statistics}. New York: John Wiley and Sons
\bibitem[\protect\citeauthoryear{Jenkins et al.}{2015}]{jenkins:2015} 
Jenkins, J.~M., Twicken, J.~D., Batalha, N.~M., et al. 2015, AJ, 150, 56
\bibitem[\protect\citeauthoryear{Karkoschka}{1994}]{karkoschka:1994} 
Karkoschka, E. 1994, Icarus, 111, 174
\bibitem[\protect\citeauthoryear{Kipping et al.}{2010}]{hatp24} 
Kipping, D.~M., Bakos, G. \'A., Hartman, J., Torres, G., Shporer, A.,
Latham, D.~W., Kov\'acs, G., Noyes, R.~W., Howard, A.~W., Fischer, D.~A.,
Johnson, J.~A., Marcy, G.~W., B\'eky, B., Perumpilly, G., Esquerdo, G.~A.,
Sasselov, D.~D., Stefanik, R.~P., L\'az\'ar, J., Papp,~I. \& S\'ari, P.,
2010, ApJ, 725, 2017
\bibitem[\protect\citeauthoryear{Kipping \& Spiegel}{2011}]{kipping:2011}
Kipping, D.~M., Spiegel, D.~S., 2011, MNRAS, 417, L88
\bibitem[\protect\citeauthoryear{Kipping et al.}{2013}]{HEK2} 
Kipping, D.~M., Hartman, J., Buchhave, L.~A., Schmitt, A.~R., Bakos, G.~\'A.
\& Nesvorn\'y, D. 2013, ApJ, 770, 101
\bibitem[\protect\citeauthoryear{Knutson et al.}{2007}]{knutson:2007}
Knutson, H.~A., Charbonneau, D., Allen, L.~E., Fortney, J.~J., Agol, E.,
Cowan, N.~B., Showman, A.~P., Cooper, C.~S. \& Megeath, S.~T., 2007,
Nature, 447, 183
\bibitem[\protect\citeauthoryear{Lehmer}{2017}]{lehmer:2017} 
Lehmer, O.~R. \& Catling, D.~C. 2017, ApJ, 845, 130
\bibitem[\protect\citeauthoryear{Loeb \& Gaudi}{2003}]{loeb:2003} 
Loeb, A. \& Gaudi, S.~B. 2003, ApJ, 588, 117
\bibitem[\protect\citeauthoryear{Lopez \& Fortney}{2014}]{lopez:2014} 
Lopez, E.~D. \& Fortney, J.~J. 2014, ApJ, 792, 1
\bibitem[\protect\citeauthoryear{Lucy \& Sweeney}{1971}]{lucy:1971} 
Lucy, L.~B. \& Sweeney, M.~A. 1971, AJ, 76, 544
\bibitem[\protect\citeauthoryear{Mallama}{2009}]{mallama:2009} 
Mallama, A. 2009, Icarus, 204, 11
\bibitem[\protect\citeauthoryear{Mathur et al.}{2017}]{mathur:2017} 
Mathur, S., Huber, D., Batalha, N.~M., et al. 2017, ApJS, 229, 30
\bibitem[\protect\citeauthoryear{Mazeh \& Faigler}{2010}]{mazeh:2010} 
Mazeh, T. \& Faigler, S. 2010, A\&A, 521, 59
\bibitem[\protect\citeauthoryear{Mazeh et al.}{2013}]{mazeh:2013} 
Mazeh, T., Nachmani, G., Holczer, T., Fabrycky, D.~C., Ford, E.~B.,
Sanchis-Ojeda, R., Sokol, G., Rowe, J.~F., Zucker, S., Agol, E.,
Carter, J.~A., Lissauer, J.~J., Quintana, E.~V., Ragozzine, D.,
Steffen, J.~H. \& Welsh, W., 2013, ApJS, 208, 16
\bibitem[\protect\citeauthoryear{Mcquillan et al.}{2014}]{mcquillan:2014} 
Mcquillan, A., Mazeh, T. \& Aigrain, S. 2014, ApJS, 211, 24
\bibitem[\protect\citeauthoryear{Morris \& Naftilan}{1993}]{morris:1993} 
Morris, S.~L. \& Naftilan, S.~A. 1993, ApJ, 419, 344
\bibitem[\protect\citeauthoryear{Rogers}{2015}]{rogers:2015} 
Rogers, L.~A. 2015, ApJ, 801, 41
\bibitem[\protect\citeauthoryear{Rouan et al.}{2011}]{rouan:2011} 
Rouan, D., Deeg, H.~J., Demangeon, O., Samuel, B., Cavarroc, C., Fegley, B. \&
L\'eger, A. 2011, ApJ, 741, 30
\bibitem[\protect\citeauthoryear{Rowe et al.}{2010}]{rowe:2010} 
Rowe, J., Borucki, W.~J., Koch, D., Howell, S.~B., Basri, G., Batalha, N.,
Brown, T.~M., Caldwell, D., Cochran, W.~D., Dunham, E., Dupree, A.~K.,
Fortney, J.~J., Gautier, T.~N.. Gilliland, R.~L., Jenkins, J., Latham, D.~W.,
Lissauer, J.~J., Marcy, G., Monet, D.~G., Sasselov, D., Welsh, W.~F.,
2010, Ap, 713, 150
\bibitem[\protect\citeauthoryear{Rowe \& Thompson}{2015}]{rowe:2015} 
Rowe, J. \& Thompson, S.~E. 2015, arXiv e-prints: 1504.00707
\bibitem[\protect\citeauthoryear{Samsing}{2015}]{samsing:2015}
Samsing, J. 2015, ApJ, 807, 65
\bibitem[\protect\citeauthoryear{Sanchis-Ojeda et al.}{2013}]{sanchis:2013}
Sanchis-Ojeda, R., Rappaport, S., Winn, J.~N., Levine, A., Kotson, M.~C.,
Latham, D.~W. \& Buchhave, L.~A. 2013, ApJ, 774, 54
\bibitem[\protect\citeauthoryear{Sandford \& Kipping}{2017}]{sandford:2017}
Sandford, E. \& Kipping, D.~M., 2017, AJ, 154, 228
\bibitem[\protect\citeauthoryear{Santerne et al.}{2011}]{santerne:2011}
Santerne, A., Bonomo, A.~S., H\'ebrard, G., Deleuil, M., Moutou, C.,
Almenara, J.-M., Bouchy, F. \& D\'iaz, R.~F. 2011, A\&A, 536, 70
\bibitem[\protect\citeauthoryear{Schwarz}{1978}]{schwarz:1978}
Schwarz, G.~E. 1978, Annals of Statistics, 6, 461
\bibitem[\protect\citeauthoryear{Seager \& Sasselov}{2000}]{seager:2000}
Seager, S. \& Sasselov, D.~D., 2000, ApJ, 537, 916
\bibitem[\protect\citeauthoryear{Seager et al.}{2000}]{seager:2000b}
Seager, S., Whitney, B.~A., \& Sasselov, D.~D.\ 2000, ApJ, 540, 504 
\bibitem[\protect\citeauthoryear{Seager \& Mall\'en-Ornelas}{2003}]{seager:2003}
Seager, S. \& Mall\'en-Ornelas, G., 2003, ApJ, 585, 1038
\bibitem[\protect\citeauthoryear{Sheets \& Deming}{2014}]{sheets:2014} 
Sheets, H.~A. \& Deming, D. 2014, ApJ, 794, 133
\bibitem[\protect\citeauthoryear{Sheets \& Deming}{2017}]{sheets:2017} 
Sheets, H.~A. \& Deming, D. 2017, AJ, 154, 160
\bibitem[\protect\citeauthoryear{Shporer et al.}{2014}]{shporer:2014} 
Shporer, A., O'Rourke, J.~G., Knutson, H.~A., Szab\'o, G.~M., Zhao, M.,
Burrows, A., Fortney, J., Agol, E., Cowan, N.~B., Desert, J.-M., Howard, A.~W.,
Isaacson, H., Lewis, N.~K., Showman, A.~P. \& Todorov, K.~O. 2014, ApJ, 788, 92
\bibitem[\protect\citeauthoryear{Silverman}{1986}]{silverman:1986} 
Silverman, B.~W., 1986, Density Estimation for Statistics and Data Analysis.
London: Chapman and Hall
\bibitem[\protect\citeauthoryear{Sliski \& Kipping}{2014}]{sliski:2014}
Sliski D.~H., Kipping D.~M., 2014, ApJ, 788, 148
\bibitem[\protect\citeauthoryear{Smith et al.}{2012}]{smith:2012} 
Smith, J.~C., Stumpe, M.~C., Van Cleve, J.~E., Jenkins, J.~M., Barclay, T.~S.,
Fanelli, M.~N., Girouard, F.~R., Kolodziejczak, J.~J., McCauliff, S.~D.,
Morris, R.~L. \& Twicken, J.~D. 2012, PASP, 124, 1000
\bibitem[\protect\citeauthoryear{Snellen et al.}{2010}]{snellen:2010}
Snellen, I.~A.~G., de Kok, R.~J., de Mooij, E.~J.~W. \& Albrecht, S., 2010,
Nature, 465, 1049
\bibitem[\protect\citeauthoryear{Spiegel et al.}{2010}]{spiegel:2010}
Spiegel, David S., Burrows, A., Ibgui, L., Hubeny, I., \& Milsom, J. A. 2010,
ApJ, 709, 149 
\bibitem[\protect\citeauthoryear{Stumpe et al.}{2012}]{stumpe:2012} 
Stumpe, M.~C., Smith, J.~C., Van Cleve, J.~E., Twicken, J.~D., Barclay, T.~S.,
Fanelli, M.~N., Girouard, F.~R., Jenkins, J.~M., Kolodziejczak, J.~J.,
McCauliff, S.~D. \& Morris, R.~L. 2012, PASP, 124, 985
\bibitem[\protect\citeauthoryear{Sudarsky et al.}{2000}]{sudarsky:2000}
Sudarsky, D., Burrows, A., Pinto, P. 2000, ApJ, 538, 885 
\bibitem[\protect\citeauthoryear{Teachey et al.}{2018}]{teachey:2017} 
Teachey, A., Kipping, D.~M. \& Schmitt, A. 2018, AJ, 155, 36
\bibitem[\protect\citeauthoryear{Thompson}{2015}]{thompson:2015}
Thompson, S. E., 2015, Kepler Data Release 24 Notes, KSCI-19064-002,
(Moffett Field: NASA Ames Research Center).
\bibitem[\protect\citeauthoryear{Winn}{2010}]{winn:2010}
Winn, J.~N., 2010, ``Transits \& Occultations'' in ``EXOPLANETS'', eds.
S. Seager, University of Arizona Press (Tucson, AZ)
\bibitem[\protect\citeauthoryear{Silverman}{1986}]{silverman:1986} Silverman, B. W., 1986, Density Estimation for Statistics and Data Analysis, Chapman \&
Hall, London.

\end{thebibliography}
